\pdfoutput=1  
\documentclass[journal]{IEEEtran}

\newif\ifreviewaids
\reviewaidsfalse

\usepackage{cite}
\usepackage{amsmath,amssymb,amsfonts}
\usepackage{algorithmic}
\usepackage{graphicx}
\usepackage{textcomp}
\usepackage{xcolor}
\usepackage[hidelinks]{hyperref}
\ifreviewaids
  \usepackage[switch]{lineno}
  \usepackage{fancyhdr}
\fi
\usepackage{booktabs}
\usepackage{multirow}
\usepackage{array}
\usepackage{subfig}
\usepackage{bm}          
\usepackage{siunitx}     
\usepackage{placeins}    
\usepackage{tikz}
\usetikzlibrary{arrows.meta,positioning,calc,fit,backgrounds,shapes.geometric}
\DeclareSIUnit{\sample}{Sa}
\DeclareSIUnit{\dBm}{dBm}
\DeclareSIUnit{\dBsm}{dBsm}

\tikzset{
  node distance = 3.6mm and 4.2mm,
  blk/.style   = {draw, rounded corners=1pt, align=center, font=\scriptsize,
                  inner sep=2pt, minimum height=5.4mm},
  phy/.style   = {blk, fill=blue!6},
  ric/.style   = {blk, fill=orange!8},
  ext/.style   = {blk, fill=black!5},
  flow/.style  = {-{Latex[length=1.6mm]}, thin},
  proto/.style = {-{Latex[length=1.6mm]}, thin, dashed, gray!70},
  lbl/.style   = {font=\scriptsize, inner sep=1pt, align=center},
  slbl/.style  = {font=\tiny, inner sep=1pt, align=center},
}


\newcommand{\CAMPflatCont}{54.2 $\pm$ 3.7}

\newcommand{\CAMPflatTrackIds}{2.6 $\pm$ 0.5}
\newcommand{\CAMPflatN}{5}

\newcommand{\CAMPflatRangeR}{8.7}
\newcommand{\CAMPflatVelR}{0.44}
\newcommand{\CAMPflatAzR}{7.0}

\newcommand{\CAMPflatContM}{54.2}

\newcommand{\CAMPflatIdSwitch}{38}

\newcommand{\CAMPflatContBridge}{88.31 $\pm$ 0.90}

\newcommand{\CAMPflatTrackIdsBridge}{3.00 $\pm$ 0.00}

\newcommand{\CAMPflatRangeRmseBridge}{6.97 $\pm$ 0.69}
\newcommand{\CAMPflatXrangeRmseBridge}{4.44 $\pm$ 0.27}
\newcommand{\CAMPflatDetGapMed}{0.6}   
\newcommand{\CAMPflatDetGapMax}{1.9}    
\newcommand{\CAMPflatCoastMaxS}{8}       

\newcommand{\CAMPflatContSim}{93.95 $\pm$ 0.03}

\newcommand{\CAMPflatContBridgeSim}{94.03 $\pm$ 0.30}

\newcommand{\CAMPflatContSimDelta}{0.1}   
\newcommand{\CAMPflatContWallDelta}{34.1} 

\newcommand{\CAMPpdCpis}{237}
\newcommand{\CAMPpdFloorPd}{0.73}

\newcommand{\CAMPpdLevels}{5}
\newcommand{\CAMPpdOpPd}{0.90}
\newcommand{\CAMPpdOpSnr}{23.1}

\newcommand{\CAMPpdRcsHi}{+20}
\newcommand{\CAMPpdRcsHiPd}{0.95}
\newcommand{\CAMPpdRcsLo}{-20}
\newcommand{\CAMPpdRcsLoPd}{0.73}
\newcommand{\CAMPpdRcsSpan}{40}
\newcommand{\CAMPpdSatSnr}{24}
\newcommand{\CAMPpdSeeds}{3}

\newcommand{\CAMPpfaAfter}{8.9\times10^{-5}}

\newcommand{\UPArows}{2}
\newcommand{\UPAcols}{4}
\newcommand{\UPAelems}{8}

\newcommand{\UPAcampHeightRmse}{\MheightC}    
\newcommand{\UPAcampHeightSD}{\MheightSDC}    

\newcommand{\UPAliveEndfireEl}{1.2}   
\newcommand{\UPAliveBroadsideEl}{15.1}

\newcommand{\MheightC}{4.5}\newcommand{\MheightSDC}{0.3}

\newcommand{\MTnTargets}{3}
\newcommand{\MTnTargetsWord}{Three}
\newcommand{\MTlatEkfMed}{0.58}    \newcommand{\MTlatEkfMax}{2.5}
\newcommand{\MTlatEtwoMed}{6.7}      \newcommand{\MTlatEtwoPnf}{9.9}
\newcommand{\MTlatTotalMed}{148}
\newcommand{\MTlatRecords}{673}
\newcommand{\MTconfig}{$1\times$UE, 8-port UPA}

\newcommand{\MTaRcs}{-8}          \newcommand{\MTaEl}{23.3}
\newcommand{\MTbRcs}{-14}         \newcommand{\MTbEl}{40.7}
\newcommand{\MTcRcs}{-20}         \newcommand{\MTcEl}{8.2}
\newcommand{\MTaAlt}{55}          \newcommand{\MTbAlt}{50}   \newcommand{\MTcAlt}{30}
\newcommand{\MTaRangeSpan}{42--61}
\newcommand{\MTbRangeSpan}{30--60}
\newcommand{\MTcRangeSpan}{35--79}

\newcommand{\MTrangeRes}{8.03}
\newcommand{\MTbsrs}{37.4}
\newcommand{\MTdopRes}{0.141}     
\newcommand{\MTdopBin}{0.0706}    
\newcommand{\MTnotch}{0.42}       

\newcommand{\MTaNotch}{21}  \newcommand{\MTbNotch}{10}  \newcommand{\MTcNotch}{8}

\newcommand{\MTpredResolvable}{95}   \newcommand{\MTrendResolvable}{96}
\newcommand{\MTpredElOnly}{59}       \newcommand{\MTrendElOnly}{58}

\newcommand{\MTrawCapOld}{64}

\newcommand{\MTrawCapDerived}{560}
\newcommand{\MTdetsCapDerived}{18}
\newcommand{\MTescSweepLo}{4}        \newcommand{\MTescSweepHi}{32}
\newcommand{\MTescSweepCovLo}{67.4}  
\newcommand{\MTescSweepIdswLo}{18.0} 
\newcommand{\MTescRateMean}{2.0}     \newcommand{\MTescRateMin}{1.7}
\newcommand{\MTescRateMax}{2.6}

\newcommand{\MTrawSatBefore}{17 of 18}
\newcommand{\MTdetsBefore}{2}

\newcommand{\MTfalseFrac}{19}

\newcommand{\MTgainUl}{115}
\newcommand{\MTgainSettle}{115.2}     
\newcommand{\MTsrsGap}{202}           
\newcommand{\MTposAuto}{21}           
\newcommand{\MTposPinned}{98}         

\newcommand{\MTtapsPerUav}{5}
\newcommand{\MTperTapRatio}{2.03}
\newcommand{\MTloadVsStable}{1.59}
\newcommand{\MTrachHealthy}{19}
\newcommand{\MTrntiHealthy}{4}
\newcommand{\MTrachStarved}{1083}
\newcommand{\MTrntiStarved}{162}

\newcommand{\MTcpiPeriod}{0.64}
\newcommand{\MTwallGap}{10.9}
\newcommand{\MTdilation}{17}
\newcommand{\MTttl}{5}
\newcommand{\MTnRxPorts}{8}
\newcommand{\MTrtBudget}{625}
\newcommand{\MTpipelineMs}{141}
\newcommand{\MTpipelineMax}{152}
\newcommand{\MTcpiWindowMs}{640}
\newcommand{\MTbatchUs}{6320}
\newcommand{\MToverBudget}{10.1}
\newcommand{\MTrateDesign}{1.56}
\newcommand{\MTrateWall}{0.09}
\newcommand{\MTaStep}{1.28}  \newcommand{\MTbStep}{1.92}  \newcommand{\MTcStep}{1.60}

\newcommand{\MTcovPooled}{41.5 $\pm$ 5.5}
\newcommand{\MTceilingWall}{38.0}   \newcommand{\MTceilingSim}{100}
\newcommand{\MTcovWall}{14.6}       \newcommand{\MTcovSim}{41.5}
\newcommand{\MTfmWall}{70.1}        \newcommand{\MTfmSim}{4.3}
\newcommand{\MTidfWall}{0.21}       \newcommand{\MTidfSim}{0.44}
\newcommand{\MTprimaryWall}{31.52}  \newcommand{\MTprimarySim}{32.33}
\newcommand{\MTmotaWall}{0 / 10 / 20}
\newcommand{\MTmotaSim}{9 / 9 / 12}
\newcommand{\MTmotaTargetRuns}{30}   
\newcommand{\MTmotaSimLost}{12}      
\newcommand{\MTrmseWall}{9.22}      \newcommand{\MTrmseSim}{9.23}
\newcommand{\MTgospaMissedWall}{22.39}  \newcommand{\MTgospaMissedSim}{18.48}
\newcommand{\MTgospaLocWall}{3.01}      \newcommand{\MTgospaLocSim}{8.43}
\newcommand{\MTgospaFalseWall}{4.18}    \newcommand{\MTgospaFalseSim}{12.11}
\newcommand{\MTidsPerTgtWall}{1.33} \newcommand{\MTidsPerTgtSim}{1.30}
      
\newcommand{\MTaCovSim}{76.5}  \newcommand{\MTbCovSim}{12.9}  \newcommand{\MTcCovSim}{18.0}

\newcommand{\MTbAzResidGround}{-10.61}  \newcommand{\MTbAzResidCone}{-0.65}
\newcommand{\MTbAzSdGround}{13.25}      \newcommand{\MTbAzSdCone}{7.04}
\newcommand{\MTsigmaAzMeas}{6.2}        \newcommand{\MTsigmaRMeas}{6.0}
\newcommand{\MTrangeBias}{7.5}
\newcommand{\MTrangeResidBefore}{-7.8 to -6.8}
\newcommand{\MTrangeResidAfter}{-0.6 to +0.7}
\newcommand{\MTelRho}{0.25}
\newcommand{\MTzSingleLook}{13--23}
\newcommand{\MTzReported}{5--16}
\newcommand{\MTaZsdBefore}{18.5}        \newcommand{\MTaZsdAfter}{7.3}
\newcommand{\MTsigmaElMeas}{9.3}

\newcommand{\MTaZbias}{-0.2}  \newcommand{\MTaZsd}{7.8}   \newcommand{\MTaZin}{100}
\newcommand{\MTbZbias}{+0.5}  \newcommand{\MTbZsd}{34.7}  \newcommand{\MTbZin}{80}
\newcommand{\MTcZbias}{+5.8}  \newcommand{\MTcZsd}{22.4}  \newcommand{\MTcZin}{74}
\newcommand{\MTaZsupp}{21}  \newcommand{\MTbZsupp}{3}  \newcommand{\MTcZsupp}{67}
\newcommand{\MTaElRatio}{2.5}  \newcommand{\MTbElRatio}{4.4}  \newcommand{\MTcElRatio}{0.9}

\newcommand{\MTaltRmse}{8.92 $\pm$ 3.23}   

\newcommand{\MTcAltRmse}{18.57 $\pm$ 8.13}
\newcommand{\MTaExcl}{82}  \newcommand{\MTbExcl}{70}  \newcommand{\MTcExcl}{56}
\newcommand{\MTcontested}{73.7 $\pm$ 4.8}
\newcommand{\MTcontestedNoEl}{87.1 $\pm$ 3.6}
\newcommand{\MTcontestedEl}{73.7 $\pm$ 4.8}
\newcommand{\MTcontestedElN}{10}     
\newcommand{\MTbaseCov}{32.4 $\pm$ 8.8}      \newcommand{\MTbaseIdfOne}{0.233 $\pm$ 0.049}
\newcommand{\MTbaseIdentifiers}{1.43 $\pm$ 0.42} \newcommand{\MTbaseIdsw}{2.0 $\pm$ 1.0}
\newcommand{\MTbaseRmse}{13.53 $\pm$ 0.75}   \newcommand{\MTbasePrimaryRmse}{38.28 $\pm$ 4.60}

\newcommand{\MTbaseCservedGtGated}{5.1 $\pm$ 5.4}
\newcommand{\MTcservedGtGated}{18.0 $\pm$ 9.1}
\newcommand{\MTrmse}{9.23 $\pm$ 1.53}        

\newcommand{\MTbirthElSep}{16}
\newcommand{\MTcovNoElSep}{41.5}   \newcommand{\MTcovElSep}{57.9}
\newcommand{\MTidsNoElSep}{1.30 $\pm$ 0.41}  \newcommand{\MTidsElSep}{3.33 $\pm$ 0.60}
\newcommand{\MTidswNoElSep}{3.0 $\pm$ 2.1}   \newcommand{\MTidswElSep}{11.2 $\pm$ 3.2}
\newcommand{\MTidfNoElSep}{0.435}  \newcommand{\MTidfElSep}{0.359}
\newcommand{\MTaPosRmse}{8.13}   \newcommand{\MTbPosRmse}{24.32}
\newcommand{\MTcPosRmse}{37.74}
\newcommand{\MTposSpread}{4.6}
\newcommand{\MTkpiLo}{1}  \newcommand{\MTkpiHi}{10}
\newcommand{\MTaPosRmseAR}{8.03}   \newcommand{\MTbPosRmseAR}{36.95}
\newcommand{\MTcPosRmseAR}{37.95}

\newcommand{\MTbContAR}{36.0}  \newcommand{\MTbCont}{21.2}
\newcommand{\MTassocForced}{21.6 $\pm$ 11.3}

\newcommand{\MTnmsRadius}{17.1}
\newcommand{\MThdgBaseNmsMerged}{12.7}  \newcommand{\MThdgDiagNmsMerged}{31.7}

\newcommand{\MTaAzMax}{22}   \newcommand{\MTbAzMax}{59}   
\newcommand{\MTaAperture}{0.92} \newcommand{\MTbAperture}{0.51} \newcommand{\MTcAperture}{0.93}
\newcommand{\MTaTotalRmse}{8.1} \newcommand{\MTbTotalRmse}{24.3} \newcommand{\MTcTotalRmse}{37.7}
\newcommand{\MTcElK}{0.78}      
\newcommand{\MTelObsK}{2}       
\newcommand{\MTbGndRange}{35}   

\newcommand{\MTelSeeds}{8}
\newcommand{\MTelBaseRate}{5--8}          
\newcommand{\MTelBaseRateLo}{4.7}  \newcommand{\MTelBaseRateHi}{7.3}
\newcommand{\MTelContestedNat}{30}        
\newcommand{\MTelEnrichLo}{4.2}  \newcommand{\MTelEnrichHi}{6.5}

\newcommand{\MTcoastDelta}{-0.03}     
\newcommand{\MTcoastP}{0.91}          
\newcommand{\MTcoastDetCov}{96}       
\newcommand{\MTcoastDetOff}{80.6}     
\newcommand{\MTcoastDetOn}{58.7}      
\newcommand{\MTescBservedOff}{12.9}  \newcommand{\MTescBservedOn}{31.4}
\newcommand{\MTescBservedP}{0.004}
\newcommand{\MTescCservedOff}{18.0}  \newcommand{\MTescCservedOn}{36.1}
\newcommand{\MTescCservedP}{0.008}
\newcommand{\MTescCovOff}{41.5}      \newcommand{\MTescCovOn}{57.9}
\newcommand{\MTescCovP}{0.002}

\newcommand{\MTescCovDelta}{16.4}

\newcommand{\MTescSigmaK}{1.72}
\newcommand{\MTpersistCovKept}{11.5}
\newcommand{\MTpersistIdswKept}{20.7}
\newcommand{\MTpersistHoldLo}{13}
\newcommand{\MTpersistHoldHi}{36}

\newcommand{\MTpermTested}{45}      
\newcommand{\MTpermFlagged}{2}      
\newcommand{\MTpermSplits}{126}     
\newcommand{\MTpermMedian}{1}       
\newcommand{\MTpermMax}{11}         
\newcommand{\MTpermP}{0.46}         
\newcommand{\MTcampBatch}{5}
\newcommand{\MTpermTestedDiag}{47}   \newcommand{\MTpermFlaggedDiag}{3}
\newcommand{\MTpermMedianDiag}{2}    \newcommand{\MTpermPDiag}{0.39}
\newcommand{\MTpCov}{0.004}        \newcommand{\MTpIdf}{0.002}
\newcommand{\MTpRmse}{0.002}       \newcommand{\MTpCserved}{0.008}
\newcommand{\MTpIdentifiers}{0.52} \newcommand{\MTpIdsw}{0.29}
\newcommand{\MTpPrimary}{0.014}

\newcommand{\MTidswBatchOld}{1.40}  \newcommand{\MTidswBatchNew}{4.60}
\newcommand{\MTidswBatchP}{0.032}   
\newcommand{\MTrmseGated}{9.23}      \newcommand{\MTbaseRmseGated}{13.53}
\newcommand{\MTrmseUngated}{35.54}   \newcommand{\MTbaseRmseUngated}{31.41}
\newcommand{\MTcovGated}{41.5}       \newcommand{\MTbaseCovGated}{32.4}
\newcommand{\MTcovUngated}{74.7}     \newcommand{\MTbaseCovUngated}{66.7}
\newcommand{\MTrmseDeltaGated}{-4.30}    \newcommand{\MTrmsePGated}{0.002}
\newcommand{\MTrmseDeltaUngated}{+4.13}  \newcommand{\MTrmsePUngated}{0.12}
     
\newcommand{\MTcampN}{10}
\newcommand{\MTaRangeRmse}{7.51 $\pm$ 3.16}    \newcommand{\MTaCrossRmse}{3.13 $\pm$ 0.43}
\newcommand{\MTbRangeRmse}{17.21 $\pm$ 5.92}   \newcommand{\MTbCrossRmse}{17.18 $\pm$ 23.17}
\newcommand{\MTcRangeRmse}{20.45 $\pm$ 7.69}   \newcommand{\MTcCrossRmse}{31.71 $\pm$ 10.33}
  \newcommand{\MTbCrossRmseMed}{5.06}
\newcommand{\MTcRangeRmseMed}{22.16}  \newcommand{\MTcCrossRmseMed}{33.84}
\newcommand{\MTbCrossRmseLo}{1.48}    \newcommand{\MTbCrossRmseHi}{67.43}
\newcommand{\MTbCrossOutliers}{2}     
\newcommand{\MTcampNb}{8}   
\newcommand{\MTcampNc}{9}   

\newcommand{\MTidentifiers}{1.30 $\pm$ 0.41}  \newcommand{\MTidsw}{3.0 $\pm$ 2.1}
\newcommand{\MTidfOne}{0.43 $\pm$ 0.04}       
\newcommand{\MTprimaryRmse}{32.33 $\pm$ 4.08}
\newcommand{\MTgospa}{25.41 $\pm$ 1.19}        \newcommand{\MTgospaLoc}{8.43 $\pm$ 1.69}
\newcommand{\MTgospaMissed}{18.48 $\pm$ 0.89}  \newcommand{\MTgospaFalse}{12.11 $\pm$ 2.14}
\newcommand{\MTaPd}{83.7 $\pm$ 2.1}  \newcommand{\MTbPd}{72.7 $\pm$ 4.0}  \newcommand{\MTcPd}{53.7 $\pm$ 5.7}
\newcommand{\MTabsentFalse}{0.04 $\pm$ 0.00}      
\newcommand{\MTovermergeIds}{3.0 $\pm$ 0.0}     

\newcommand{\MThdgBaseNotch}{9.5}  \newcommand{\MThdgDiagNotch}{15.9}
\newcommand{\MThdgBaseReport}{90.5} \newcommand{\MThdgDiagReport}{84.1}
\newcommand{\MThdgBaseAmb}{0.0}    \newcommand{\MThdgDiagAmb}{74.6}
\newcommand{\MThdgDiagAmbRange}{44.4}   
\newcommand{\MThdgBaseDaz}{29.5}   \newcommand{\MThdgDiagDaz}{1.6}
\newcommand{\MThdgBaseDel}{17.5}   \newcommand{\MThdgDiagDel}{4.5}
\newcommand{\MThdgBaseElDec}{12.7} \newcommand{\MThdgDiagElDec}{1.6}
\newcommand{\MThdgDiagIdfOne}{0.42 $\pm$ 0.05}   \newcommand{\MThdgDiagIdentifiers}{2.30 $\pm$ 0.31}
\newcommand{\MThdgDiagIdsw}{8.6 $\pm$ 2.9}     \newcommand{\MThdgDiagGospa}{24.59 $\pm$ 1.31}
      
      \newcommand{\MThdgDiagContested}{97.4 $\pm$ 1.4}
\newcommand{\MThdgDiagCampN}{10}
      \newcommand{\MTelAmbPct}{34.7}

\newcommand{\MTelTrueC}{8.12}   \newcommand{\MTelBiasC}{-0.75}  
\newcommand{\MTelTrueA}{23.18}  \newcommand{\MTelBiasA}{-1.01}  
\newcommand{\MTelTrueB}{34.61}  \newcommand{\MTelBiasB}{+0.59}  
\newcommand{\MTelBiasSpread}{1.6}    
\newcommand{\MTelSlope}{0.046}  \newcommand{\MTelSlopeLo}{0.011}  \newcommand{\MTelSlopeHi}{0.083}
\newcommand{\MTelSpanLo}{7.9}   \newcommand{\MTelSpanHi}{40.9}    \newcommand{\MTelSlopeSwing}{1.50}
\newcommand{\MTelRmseClean}{3.74}    \newcommand{\MTelCalConst}{3.68}
\newcommand{\MTelCalLin}{3.67}       \newcommand{\MTelCalTgt}{3.65}
\newcommand{\MTelCalBestPct}{2.4}
\newcommand{\MTelAmbBiasC}{+22.01}   \newcommand{\MTelAmbSdC}{16.49}
\newcommand{\MTelRmseAmb}{23.19}
\newcommand{\MTelPriorArmLo}{+4.4}   \newcommand{\MTelPriorArmHi}{+5.0}
\newcommand{\MTelPriorXLo}{-3.9}     \newcommand{\MTelPriorXHi}{-7.8}

\newcommand{\MThcContestedBase}{73.74}\newcommand{\MThcContestedDiag}{97.45}\newcommand{\MThcContestedT}{14.3}   
\newcommand{\MThcPdABase}{83.71}\newcommand{\MThcPdADiag}{77.93}\newcommand{\MThcPdAT}{3.5}   
\newcommand{\MThcPdBBase}{72.69}\newcommand{\MThcPdBDiag}{65.25}\newcommand{\MThcPdBT}{4.5}   
\newcommand{\MThcPdCBase}{53.73}\newcommand{\MThcPdCDiag}{46.94}\newcommand{\MThcPdCT}{2.4}   
\newcommand{\MThcIdfBase}{0.43}\newcommand{\MThcIdfDiag}{0.42}\newcommand{\MThcIdfT}{0.8}   
\newcommand{\MThcGospaBase}{25.41}\newcommand{\MThcGospaDiag}{24.59}\newcommand{\MThcGospaT}{1.4}   
\newcommand{\MThcIdsBase}{1.30}\newcommand{\MThcIdsDiag}{2.30}\newcommand{\MThcIdsT}{5.8}   
\newcommand{\MThcIdswBase}{3.00}\newcommand{\MThcIdswDiag}{8.60}\newcommand{\MThcIdswT}{4.7}   
\newcommand{\MThcPdADrop}{5.8}  \newcommand{\MThcPdBDrop}{7.4}  \newcommand{\MThcPdCDrop}{6.8}

\newcommand{\MTjpdaStraightN}{10}
\newcommand{\MTjpdaStraightCovPda}{41.54}
\newcommand{\MTjpdaStraightCovJpda}{49.89}

\newcommand{\MTjpdaStraightCovWins}{8}

\newcommand{\MTjpdaStraightCovP}{0.027}

\newcommand{\MTjpdaStraightIdswPda}{3.00}
\newcommand{\MTjpdaStraightIdswJpda}{2.30}

\newcommand{\MTjpdaStraightIdswP}{0.656}
\newcommand{\MTjpdaStraightIdsPda}{1.30}
\newcommand{\MTjpdaStraightIdsJpda}{1.40}

\newcommand{\MTjpdaStraightIdsP}{0.719}

\newcommand{\MTjpdaStraightGospaP}{1.000}

\newcommand{\MTjpdaStraightGospaLocPda}{8.43}
\newcommand{\MTjpdaStraightGospaLocJpda}{11.06}

\newcommand{\MTjpdaStraightGospaLocLosses}{10}
\newcommand{\MTjpdaStraightGospaLocP}{0.002}

\newcommand{\MTjpdaDiagN}{10}

\newcommand{\MTjpdaDiagCovP}{0.492}

\newcommand{\MTjpdaDiagIdsPda}{2.30}
\newcommand{\MTjpdaDiagIdsJpda}{1.97}

\newcommand{\MTjpdaDiagIdsP}{0.039}
\newcommand{\MTjpdaDiagRmsePda}{9.50}
\newcommand{\MTjpdaDiagRmseJpda}{8.46}

\newcommand{\MTjpdaDiagGospaPda}{24.59}
\newcommand{\MTjpdaDiagGospaJpda}{23.58}

\newcommand{\MTjpdaDiagGospaWins}{10}

\newcommand{\MTjpdaDiagGospaP}{0.002}

\newcommand{\MTjpdaDiagGospaFalsePda}{11.04}
\newcommand{\MTjpdaDiagGospaFalseJpda}{9.48}

\newcommand{\MTjpdaDiagGospaFalseP}{0.006}

\newcommand{\MTjpdaStraightLocSameCard}{1.04}
\newcommand{\MTjpdaStraightLocDiffCard}{7.27}
\newcommand{\MTjpdaStraightMissDiffCard}{-6.29}
\newcommand{\MTjpdaStraightSameCardPct}{73.2}

\newcommand{\MTbgStraightN}{10}
\newcommand{\MTbgStraightCovKtwo}{41.54}
\newcommand{\MTbgStraightCovKfour}{39.37}

\newcommand{\MTbgStraightCovP}{0.062}

\newcommand{\MTbgStraightIdsKtwo}{1.30}
\newcommand{\MTbgStraightIdsKfour}{1.00}

\newcommand{\MTbgStraightIdsP}{0.062}

\newcommand{\MTbgStraightGospaFalseKtwo}{12.11}
\newcommand{\MTbgStraightGospaFalseKfour}{8.52}

\newcommand{\MTbgStraightGospaFalseP}{0.004}

\newcommand{\MTbgDiagCovKtwo}{51.16}
\newcommand{\MTbgDiagCovKfour}{48.18}

\newcommand{\MTbgDiagCovP}{0.016}

\newcommand{\MTbgDiagIdsKtwo}{2.30}
\newcommand{\MTbgDiagIdsKfour}{1.90}

\newcommand{\MTbgDiagIdsP}{0.016}

\newcommand{\MTbgDiagGospaFalseKtwo}{11.04}
\newcommand{\MTbgDiagGospaFalseKfour}{8.12}

\newcommand{\MTbgDiagGospaFalseP}{0.002}

\newcommand{\MTbgStraightDropsConf}{0.00}

\newcommand{\MTbgDiagDropsConf}{0.00}

\ifreviewaids
  \linenumbers
\fi

\begin{document}

\title{Multi-UAV Tracking Evaluation Using 5G Uplink Signals on an O-RAN ISAC
  Simulation Testbed}

\author{Arun~K.~Gurung,~\IEEEmembership{Senior Member,~IEEE,}
  and~Satha~K.~Sathananthan%
  \thanks{A. K. Gurung is an independent consultant, Melbourne,
    Australia (e-mail: akgurung@ieee.org).}%
  \thanks{S. K. Sathananthan is with NexVis, Melbourne, Australia
    (e-mail: satha@nexvis.com.au).}}

\maketitle

\begin{abstract}
  We evaluate multi-target detection, association and
  tracking end to end on an O-RAN \emph{simulation} testbed built
  from OpenAirInterface, FlexRIC and Sionna RT that repurposes the 5G NR uplink
  sounding reference signal as a passive radar waveform, and against what a
  counter-UAS command-and-control (C2) consumer requires rather than by detection
  alone.
  \MTnTargetsWord{} UAVs differing in altitude, velocity and radar cross section
  ($\MTaRcs$ to $\MTcRcs$~dBsm) fly one bistatic pair with an
  \MTnRxPorts-element planar receive array. Once every target is detected the
  binding limit is
  \emph{contention}, not sensitivity: two targets share one nearest detection in
  \MTcontested\% of coherent processing intervals, and targets are detected far
  more often than they
  are tracked. Elevation from that array cannot separate
  targets sharing a range--Doppler cell, but it decides association in
  \MTrendElOnly\% of intervals. Concurrent tracks are exported from the
  RAN Intelligent Controller (RIC) xApp to a C2 fusion node over a
  SAPIENT interface carrying a calibrated detection
  confidence, validated at schema level against a mock fusion node.
  The evaluation is emulation-only on one geometry with \MTcampN{} noise seeds,
  the mechanisms are characterized and analyzed to identify the key factors
  influencing multi-target tracking performance.
\end{abstract}


\section{Introduction}
\label{sec:intro}

Low-altitude unmanned aerial vehicles (UAVs) in civil airspace have created a
need for scalable, low-cost counter unmanned-aircraft-system (counter-UAS)
sensing.
Dedicated radar requires spectrum, infrastructure and operational budgets that
many deployments cannot justify; cellular networks already blanket terrestrial
environments and can be reused for dual-purpose sensing, the Integrated Sensing
and Communication (ISAC) vision under study in 3GPP SA1 and
RAN~\cite{3gpp_tr22837, 3gpp_tr38867}. As of August~2026, 3GPP has
\emph{service-requirement} study material for ISAC (TR~22.837, a study
report~\cite{3gpp_tr22837}) and \emph{channel-model} work, with a Release-19
change request approved to introduce an ISAC channel model into
TR~38.901~\cite{3gpp_cr38901_isac}; TR~38.867~\cite{3gpp_tr38867} likewise
remains a study. The \SIrange{1}{10}{\metre} accuracy figures we compare against in
Section~\ref{sec:assoc:birth} are use-case targets drawn from a study report.

Open Radio Access Network (O-RAN) architecture disaggregates the base station and
exposes standardized interfaces (E2, O1, A1) through which third-party
applications (xApps/rApps) observe and control the radio stack in near-real time
\cite{oran_wg1_arch}. That makes it a natural host for ISAC processing: sensing
can run as an xApp in the Near-RT RIC, co-located with the RAN and fed detections
over the E2 application protocol~\cite{oran_wg3_e2ap}. O-RAN specifies no sensing service
model, so the detections here travel over standard E2AP carrying a
\emph{custom, experimental} service model of our own definition (SM-SENS,
Section~\ref{sec:xapp}).

Despite growing interest in 5G ISAC \cite{liu_survey_isac, zhang_dual_function},
open end-to-end testbeds combining a real NR stack, a standards-compliant RIC and
a ray-traced channel including integration to a C2 system remain scarce.
This paper builds on a validated uplink sounding reference signal (UL-SRS) sensing testbed --- a PHY-layer sensing
stage in the OpenAirInterface (OAI) gNB (rank-$K_c$ clutter-subspace deflation,
2-D Range-Doppler processing, order-statistic constant-false-alarm-rate
detection (OS-CFAR), closed-form interferometric azimuth) feeding an extended
Kalman filter (EKF) xApp
over a custom E2 sensing service model (SM-SENS), all containerized over OAI,
FlexRIC and Sionna RT.

\emph{Relationship to the companion manuscript.} The companion ~\cite{gurung_isac_testbed} owns the platform: the coordinate frame and measurement
model, the single-target detector characterization on a uniform \emph{linear}
receive array, and the removal of that reference's one structural gap --- a single
bistatic pair with a linear array has no elevation observability, so the tracker
depends on an assumed height prior --- by two independent routes, a \emph{planar}
array and a two-user \emph{multistatic} geometry, both still
single-target. This paper takes the elevation-capable
array as given and asks what happens when more than one target is present at
once; it owns the multi-target evaluation and its export to a C2 consumer. This paper is therefore \emph{self-contained for its own claims but not for its background}. The \emph{single-target reference} the multi-target results are
read against is reported here with its numbers (Section~\ref{sec:results}); deferred
are its full characterization --- range-Doppler maps, radar cross-section (RCS)
and echo-SNR sweeps,
the height-prior ablation --- and the derivations of the coordinate frame and
measurement model.

\textbf{Contributions.} This paper is an \emph{end-to-end evaluation of
multi-target UAV tracking fed to a counter-UAS command-and-control node}. A
sensing evaluation asks whether the target was detected and how accurately it was
placed; a C2 consumer additionally requires that the track keep one identity,
arrive often enough to act on, and carry a confidence it can fuse. Those
requirements select different metrics, and they turn several results that read as
adequate under the first standard into failures under the second.

\begin{enumerate}
  \item \textbf{The binding constraint is contention, not sensitivity.}
        \MTnTargets{} simultaneous UAVs, heterogeneous in altitude, velocity and radar
        cross section ($\MTaRcs$ to $\MTcRcs$~dBsm), are all detected, with per-target
        availability \MTaPd\%, \MTbPd\% and \MTcPd\% under a one-to-one assignment.
        Yet in \MTcontested\% of coherent processing intervals two targets' nearest
        detection is the \emph{same} detection, so targets are detected far more often
        than they are tracked (availability \MTbPd\% against \MTbCovSim\% coverage for
        UAV-B). The deficit is association-side, not detection starvation
        (Section~\ref{sec:assoc:availability}).

  \item \textbf{Sensor limits and evaluation limits are separable.} A digital twin that cannot run in real time
        changes the result rather than merely delaying it.
        Two constants in the evaluation path carry wall-clock units; they impose a
        coverage ceiling of \MTceilingWall\% under which ``mostly tracked'' is
        unreachable by construction, and re-scoring the same detections on the sensor's
        own timeline moves coverage \MTcovWall\% to \MTcovSim\% and inverts the
        generalized optimal sub-pattern assignment (GOSPA)
        decomposition. Three rules avoid the error (Section~\ref{sec:cadence}).

  \item \textbf{On a single array, elevation buys association rather than
          localization.} It does \emph{not} deliver three-dimensional localization for
        every target --- altitude is geometry-gated, published for one well-sited target
        and withheld by the observability rule for much of the other two, and the pooled
        altitude carries a wide interval. It \emph{does} decide association, being the
        deciding discriminant in \MTrendElOnly\% of intervals.

  \item \textbf{A standards-based multi-track export to counter-UAS C2.} Concurrent tracks are mapped to a
        standards-based sensor-to-C2 message set with a calibrated detection-confidence
        model, validated at schema level against a mock fusion node
        (Section~\ref{sec:c2}).
\end{enumerate}

\textbf{Scope and limitations.} These are evaluation results from an emulation. \emph{(i)} No result
here is measured on radios: the channel is ray-traced and convolved into
\texttt{rfsimulator}'s baseband loopback, with no RF front end, no oscillator or
timing impairment and no external interference. \emph{(ii)} The statistical
design is \MTcampN{} noise seeds on \emph{one} deterministic trajectory and
\emph{one} gNB siting, with a single alternative heading as the only geometric
variation, so the \emph{mechanisms} are what we expect to transfer
(Section~\ref{sec:assoc:results}). \emph{(iii)} The C2 adapter is exercised
against an in-repository mock fusion node. \emph{(iv)} Identifier stability remains
open --- duplicate identifiers are held concurrently on one target once the
cadence artifact is removed --- and the track-continuity prerequisite is
established for a single target, not for a multi-target feed. \emph{(v)} The
platform itself is the companion manuscript's~\cite{gurung_isac_testbed}; what
this paper needs in order to stand on its own is summarized in
Appendix~\ref{app:standalone}.

\begin{figure*}[!tbp]
  \centering
  \resizebox{\textwidth}{!}{%
    \begin{tikzpicture}
      \node[ext] (scene)
      {Scene: \MTnTargets{} UAVs\\[-1pt]\scriptsize RCS $\MTaRcs$ to $\MTcRcs$~dBsm\\[-1pt]\scriptsize differing altitude, bearing};
      \node[phy, right=12mm of scene] (ue)
      {OAI nrUE\\[-1pt]\scriptsize UL-SRS Tx\\[-1pt]\scriptsize \emph{one} bistatic pair};
      \node[ext, right=12mm of ue] (emu)
      {Channel emulator\\[-1pt]\scriptsize echo injection\\[-1pt]\scriptsize calibrated AWGN};
      \node[ext, above=7mm of emu] (rt) {Sionna RT\\[-1pt]\scriptsize ray-traced paths};

      \node[phy, below=17mm of ue] (gnb)
      {OAI gNB PHY\\[-1pt]\scriptsize ECA $\cdot$ RD map\\[-1pt]\scriptsize OS-CFAR $\cdot$ az\,+\,el};
      \node[ric, right=15mm of gnb] (ric) {Near-RT RIC\\[-1pt]\scriptsize FlexRIC};
      \node[ric, right=12mm of ric] (xapp)
      {EKF xApp\\[-1pt]\scriptsize PDA association\\[-1pt]\scriptsize 3-D filter};
      \node[ric, right=12mm of xapp] (adapt) {SAPIENT ASM\\[-1pt]\scriptsize adapter};
      \node[ext, right=15mm of adapt] (c2) {HLDMM\\[-1pt]\scriptsize fusion node (mock)};

      \draw[flow] (scene) -- node[lbl, above] {echo} (ue);
      \draw[flow] (ue)  -- node[lbl, above] {SRS} (emu);
      \draw[flow] (rt)  -- node[lbl, right] {paths} (emu);
      \draw[flow] (emu.south) -- ++(0,-6mm)
        node[lbl, above right, inner sep=1pt] {IQ} -| (gnb.north);
      \draw[flow] (gnb) -- node[lbl, above] {E2} (ric);
      \draw[flow] (ric) -- (xapp);
      \draw[flow] (xapp) -- node[lbl, above] {tracks} (adapt);
      \draw[proto] (adapt) -- node[lbl, above] {SAPIENT} (c2);

      \node[lbl, below=9mm of gnb, align=center, draw, rounded corners=1pt,
        fill=red!5, inner sep=2.5pt] (fdet)
      {\textbf{1. capacity}\\[-1pt]
        caps derived from $N_T$\\[-1pt]
        (\S\ref{sec:mt:rawcap})};
      \node[lbl, below=9mm of xapp, align=center, draw, rounded corners=1pt,
        fill=red!5, inner sep=2.5pt] (fass)
      {\textbf{2. contention}\\[-1pt]
        not sensitivity\\[-1pt]
        (\S\ref{sec:assoc:availability})};
      \node[lbl, below=9mm of adapt, align=center, draw, rounded corners=1pt,
        fill=red!5, inner sep=2.5pt] (fid)
      {\textbf{3. identity}\\[-1pt]
        not detection\\[-1pt]
        (\S\ref{sec:c2_validation})};
      \draw[flow, red!45] (fdet) -- (gnb);
      \draw[flow, red!45] (fass) -- (xapp);
      \draw[flow, red!45] (fid)  -- (adapt);

      \begin{scope}[on background layer]
        \node[draw, dashed, black!35, rounded corners=2pt,
          fit=(rt)(ue)(emu)(gnb)(ric)(xapp)(adapt)(c2),
          inner sep=2.6mm, label={[slbl, black!55]above:
              single container stack --- emulation only, no radios}] {};
      \end{scope}
    \end{tikzpicture}}

  \caption{\textbf{The evaluated pipeline.} \MTnTargets{} UAVs over a single bistatic pair with a planar receive array, from ray-traced propagation to the SAPIENT export; the chain reads left to right across two rows. Block labels: ECA, clutter-subspace deflation; RD, range--Doppler map; PDA, probabilistic data association; ASM, SAPIENT autonomous sensor module; HLDMM, fusion node (Section~\ref{sec:c2}). Red markers number the three multi-target failure points in the order the paper treats them. Emulation only; no result is measured on radios.}
  \label{fig:pipeline}
\end{figure*}
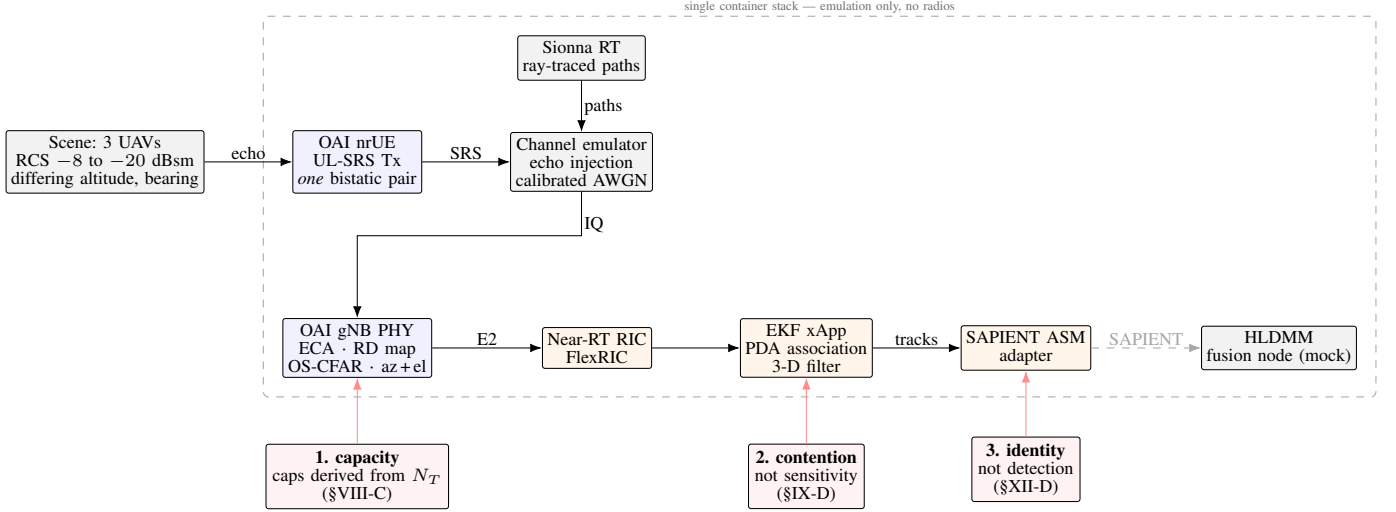

Section~\ref{sec:related} provides
cellular-ISAC literature status. Section~\ref{sec:testbed} describes the testbed and
Section~\ref{sec:sensing} the UL-SRS sensing pipeline; the coordinate system,
bistatic geometry and measurement model are developed in full in the
companion~\cite{gurung_isac_testbed}, and the configuration evaluated here is the
pipeline of Fig.~\ref{fig:pipeline}. Section~\ref{sec:results} gives the
single-target reference, Section~\ref{sec:upa} the planar-array elevation upgrade
used throughout, and Section~\ref{sec:multistatic} the two-user geometry as
context. Sections~\ref{sec:multitarget}--\ref{sec:association} present the
multi-target scenario, detector behavior and association results;
Section~\ref{sec:measmodel} re-derives the measurement model on that geometry,
Section~\ref{sec:cadence} separates sensor limits from evaluation limits, and
Section~\ref{sec:c2} describes the C2 export. Section~\ref{sec:future} sets out
the limitations and Section~\ref{sec:conclusion} concludes.

\section{Related Work}
\label{sec:related}

\emph{Multi-target tracking on communication hardware.} Bauhofer
et al.~\cite{bauhofer_mtt_isac} demonstrate multi-target tracking on a
5G-compliant ISAC proof-of-concept in a factory environment, running a
probability-hypothesis-density filter in range and radial speed against
pedestrian-like targets produced by a radar target emulator, and report
sub-\SI{1.5}{\metre} mean absolute ranging error with detection rates above
\SI{91}{\percent}. Saur et al.~\cite{saur_reliable_uav} detect small UAVs with
commercial 5G equipment operated as a \emph{monostatic} OFDM radar, at sub-meter
accuracy beyond \SI{500}{\metre} in clutter-rich surroundings.

\emph{Uplink and passive UAV sensing.} Huang et al.~\cite{huang_fuse_then_detect}
have several user equipments transmit SRS pilots while the base station receives
the UAV-scattered echoes, accumulating evidence across transmitters before
detection --- the same uplink-SRS premise this paper uses ---
and reach a \SI{4.84}{\metre} median three-dimensional position error.
Varshney et al.~\cite{varshney_multitrp_uav} study multi-TRP UAV detection in a
3GPP 5G-Advanced ISAC network at system level in simulation, and Sagduyu et
al.~\cite{sagduyu_multiscout} track moving targets multistatically from the
\emph{downlink} positioning reference signal rather than the uplink.
Commercially, AT\&T and Ericsson announced detection and tracking of multiple
drones from Massive-MIMO radios on existing towers in July
2026~\cite{att_ericsson_drone}.

\emph{Sensing in O-RAN.} Programmable real-time sensing at the O-RAN edge is an
established line of work, though largely an architectural one: dApps as a
user-plane real-time extension with I/Q exposure to the
edge~\cite{polese_dapp, polese_dapps_6gr}, proposed native ISAC support in the
architecture~\cite{oran_isac_dapp}, and the nGRG requirements~\cite{ngrg_dapp}.
These establish where a sensing function belongs and how it is fed rather than
reporting sensing results from one.

This paper's contribution is an evaluation rather than a detection, in two parts. First, multi-target tracking assessed end-to-end against what a
counter-UAS C2 consumer requires rather than against detection accuracy alone,
which changes the verdict: it makes identity, not sensitivity, the binding
constraint, and it is why a mechanism that raises coverage substantially is
disabled here (Section~\ref{sec:association}). Second, the evaluation discipline
that assessment needs --- separating sensor-timeline from wall-clock limits (Section~\ref{sec:cadence}), and distinguishing
sensitivity, contention, availability and association quality as separate failure
modes (Sections~\ref{sec:multitarget}--\ref{sec:association}). The export itself
(SAPIENT / BSI~Flex~335 / STANREC~4869, from an O-RAN RIC xApp,
Section~\ref{sec:c2}) is the element we have not found reported for a
cellular-ISAC sensor; we report it as a design validated at schema level against
a mock fusion node, not as an interoperability result.

\section{Testbed Architecture}
\label{sec:testbed}

\subsection{System Model}

The system model --- the coordinate frame, the bistatic geometry and the
measurement model --- is developed in full in the companion
paper~\cite{gurung_isac_testbed}. In brief, and only
as far as the results below require: a single gNB with a planar receive array
observes the echo of an SRS transmitted by an nrUE, and measures the delay of
that echo \emph{relative to the direct path}, so the per-pair observable is the
excess bistatic path $\Delta R_{\mathrm{bi}} = R_T + R_R - L$, from which the
tracker restores the total path by adding the known baseline $L$. Azimuth is the
angle of arrival at the gNB's own array, measured from the $+\hat{\mathbf{x}}$
boresight towards $+\hat{\mathbf{y}}$; the planar array adds an elevation angle
in the orthogonal vertical plane, extending the per-pair measurement to
$[R, v, \theta_{\mathrm{az}}, \theta_{\mathrm{el}}]$.

\subsection{Overview}

The ordering of the three failure points annotated in Fig.~\ref{fig:pipeline}: capacity first, then
contention, then identity, each becoming visible only once the previous one is
removed. One further property of the figure carries into every result below --- the
sensor cadence it marks (\SI{\MTrateDesign}{\hertz} by design) is not the rate
the emulator sustains (\SI{\MTrateWall}{\hertz}), and
Section~\ref{sec:cadence} is about what that difference does to a metric.

Fig.~\ref{fig:pipeline} shows the pipeline these results are obtained on, from
the ray-traced scene through to the counter-UAS C2 export, together with the feed
a mock fusion node received: an OAI gNB carrying the ISAC PHY sensing stage, an
OAI nrUE as the SRS transmitter, FlexRIC~\cite{schmidt_flexric} Near-RT RIC
hosting the ISAC EKF xApp, a
Sionna RT channel emulator, and a real-time dashboard. All run as containers in a
single stack, so a run is reproducible from the configuration of
Table~\ref{tab:params}.


\subsection{Radio Stack and Channel Emulation}
The radio stack and the ray-traced channel are inherited unchanged from the
single-target testbed, which describes both in
detail~\cite{gurung_isac_testbed}. In brief: the gNB is
OpenAirInterface~5G~New~Radio~(NR)~\cite{kaltenberger_oai} in monolithic
central-unit/distributed-unit (CU+DU) mode on Band
n78, with the frame structure and antenna configuration of
Table~\ref{tab:params}; a Sionna RT~\cite{hoydis_sionna} server computes, once per
simulation step, a ray-traced channel impulse response between gNB and nrUE
together with per-UAV bistatic echo parameters, which a channel-emulator container
convolves into the uplink waveform. Uplink and downlink share one ray trace by TDD
reciprocity.

Two properties of that arrangement matter for a multi-target scene specifically.
First, the ray tracer returns a separate echo per target, so the emulator injects
\emph{separate target echo paths before common receiver processing} --- $N_T$
the number of targets in the scene.
Superposition, and any consequent mutual masking, therefore happens where it
happens in a real receiver: in the sampled waveform and in the range--Doppler
map. Second, because the echo amplitude is
whatever the mesh geometry produces, absolute radar cross-section is \emph{not} a
free parameter of the scene and must be imposed afterwards --- the procedure, and
why mesh radius is not a usable RCS knob, is set out in
Section~\ref{sec:multitarget}.

\emph{Scope of the target model}
Injecting independent per-target paths is not the same as modeling a target, and
the distinction bounds several results below. Each UAV is a \emph{single diffuse
sphere}: one point-like scattering center with an aspect-dependent bistatic
return supplied by the ray tracer. The model therefore includes bistatic
geometry, aspect dependence through the rendered mesh, per-target Doppler from
platform motion, ground-bounce multipath from the rendered floor, and coherent
superposition of the injected paths at the receiver. It excludes
\emph{extended-body scattering} (a real airframe is several scattering centers
whose relative phases sweep with aspect), \emph{rotor micro-Doppler} (the
spectral signature counter-UAS classifiers most often exploit),
\emph{inter-target occlusion and shadowing}, polarization, and any
target-to-target coupling.

\section{UL-SRS Sensing Pipeline}
\label{sec:sensing}

The sensing chain is inherited unchanged from the companion
testbed~\cite{gurung_isac_testbed}.
This section states only what a reader needs to follow the multi-target results:
the four constants the resolution arguments rest on, and the three stages whose
behavior changes when more than one target is present.

The pipeline runs in two stages: the \emph{gNB PHY sensing module} (C, FFTW, once
per coherent processing interval, CPI) and the \emph{ISAC xApp EKF tracker} (Python, on the Near-RT RIC per
SM-SENS indication), a split that keeps signal processing close to the PHY while
sensing logic evolves as an xApp. UL-SRS is configured at
$T_{\mathrm{SRS}} = \SI{10}{\milli\second}$, and a CPI
spans $N_{\mathrm{occ}} = 64$ bursts, so
$T_{\mathrm{CPI}} = \SI{640}{\milli\second}$ and the bistatic Doppler resolution
is $\lambda/T_{\mathrm{CPI}} \approx \SI{0.14}{\metre\per\second}$ at
$f_c = \SI{3319.68}{\mega\hertz}$. The SRS occupies a comb-2 aperture of
$B = \SI{\MTbsrs}{\mega\hertz}$ inside the \SI{38.16}{\mega\hertz} channel ---
the aperture is what sets resolution --- so the
range resolution is $c/B = \SI{\MTrangeRes}{\metre}$. Those two cells are the yardstick
every separability claim in Section~\ref{sec:multitarget} is judged against.

\subsection{Clutter Deflation}
\label{sec:clutter}

Per receive port, the channel estimates of a CPI are collected into
$\mathbf{H} \in \mathbb{C}^{N_{\mathrm{occ}} \times K}$ over slow-time occasions
and packed subcarriers. The direct path and the static ground return are constant
across slow time after carrier-frequency-offset compensation, so they occupy the
dominant left singular subspace, and the detector removes them by deflating the
leading $K_c$ directions by power iteration, on the current CPI only with no
memory across CPIs. This is the subspace form of the disturbance-removal stage
standard in passive bistatic radar~\cite{colone_eca_passive_radar}, applied per
receive port to the SRS channel estimates rather than to a reference channel.

The \emph{rank must be held at} $K_c = 1$ in a flat-LOS scene. The UAV echo is the
second-strongest return, so $K_c \ge 2$ subtracts the target itself and leaves a
smeared residual, as the companion's rank sweep
measures~\cite{gurung_isac_testbed}. Scenes with strong static multipath require
a higher, scene-dependent rank, which this evaluation does not exercise. The
consequence for a multi-target scene is that the clutter residual --- not the
target count --- is what fills the detector's crosser list
(Section~\ref{sec:mt:rawcap}).

\subsection{Detection with Two Targets in One Cell}
\label{sec:rdmap}

Range-Doppler maps are formed per receive antenna by zero-padded transforms
(range 4096, Doppler 128 points, the range transform an \emph{inverse} DFT) and
combined non-coherently in power.
OS-CFAR~\cite{rohling_oscfar} at a $75^{\mathrm{th}}$-percentile training statistic
is applied at a nominal $P_{\mathrm{FA}} = 10^{-4}$, followed by non-maximum
suppression that keeps the locally strongest cell and discards weaker cells
within a fixed radius, so one physical target yields one detection per CPI.

Two design choices in that stage govern the multi-target results. First, the crosser list feeding suppression and the cap on
reported detections are \emph{separate} fixed-size budgets, both of which the
single-target configuration sized as constants; Section~\ref{sec:mt:rawcap}
derives both from a declared target count and shows why the first is a clutter
budget rather than a target budget. Second, CFAR blanks a guard band of Doppler
bins around zero to suppress the clutter ridge, so a target inside that notch is
not reported at all --- which makes \emph{detectability}, not resolution, the
binding constraint on scenario design (Section~\ref{sec:mt:geometry}).

\subsection{Angle Estimation}
\label{sec:aoa}

Let $\mathbf{a}$ collect the complex range-Doppler cell values of a detected
target across the receive ports. For a single dominant scatterer on a
half-wavelength array the inter-element phase step is
$\pi\sin\theta_{\mathrm{az}}$, and averaging it over the available baselines
gives the bearing in closed form,
\begin{equation}
  \hat{\theta}_{\mathrm{az}} = \arcsin\left(\frac{1}{\pi}
  \arg\left( \sum_{p} a_{p+1}\, a_p^{*} \right)\right),
  \label{eq:aoa}
\end{equation}
a single-snapshot interferometric estimator which, for one source on a uniform
linear array, coincides with the least-squares ESPRIT solution over
maximally-overlapping subarrays~\cite{roy_esprit}. The planar array applies the same ramp along the vertical
baseline to give elevation (Section~\ref{sec:upa}).

Two properties matter downstream. The estimator is approximately unbiased at high
SNR near broadside and its bias grows toward endfire, which is why the one target
placed off-broadside is the one whose cross-range error is largest
(Section~\ref{sec:assoc:birth}). And it is invoked \emph{once per detected cell}
on a rank-one snapshot, so two targets sharing a range-Doppler cell return one
detection carrying one blended bearing --- the reason a vertical aperture buys
association rather than resolution (Section~\ref{sec:mt:elevation}).

\subsection{EKF xApp Integration}
\label{sec:xapp}
Detections are assembled into a per-CPI report and delivered to the Near-RT RIC
over a custom E2 sensing service model (SM-SENS); the schema, transport and xApp
internals are the companion's~\cite{gurung_isac_testbed}. On receipt the xApp
restores the baseline, $\hat{R}_{\mathrm{bi}} = \Delta\hat{R}_{\mathrm{bi}} + L$,
converting reported excess range into the path length the observation function
expects (Section~\ref{sec:measmodel}), then runs association and one EKF
predict--update cycle per track per CPI.

The single-target configuration initiates a track from any ungated detection and
coasts five missed CPIs before deletion --- adequate for one target and actively
harmful for several, since every CPI carrying more than one detection can seed a
spurious track, the direct cause of the \CAMPflatTrackIds{} identifiers per run on
the single-target reference (Section~\ref{sec:baseline}). Replacing it is a
substantial part of the multi-target contribution
(Section~\ref{sec:association}).

\section{Evaluation and Results}
\label{sec:results}

\subsection{Simulation Setup}

\begin{table}[!tbp]
  \centering
  \caption{Radio and Sensing Configuration}
  \label{tab:params}
  \footnotesize
  \begin{tabular}{@{}p{0.40\linewidth}p{0.52\linewidth}@{}}
    \toprule
    \textbf{Parameter}                           & \textbf{Value}                                                                                                                                                                                                                                                                                                                                                                                                                                     \\
    \midrule
    NR Band (carrier)                            & n78, $f_c = \SI{3319.68}{\mega\hertz}$                                                                                                                                                                                                                                                                                                                                                                                                             \\
    Subcarrier spacing                           & 30~kHz (numerology~1)                                                                                                                                                                                                                                                                                                                                                                                                                              \\
    Channel bandwidth                            & \SI{40}{\mega\hertz} nominal; 106~PRBs $=$ \SI{38.16}{\mega\hertz} occupied                                                                                                                                                                                                                                                                                                                                                                        \\
    TDD pattern                                  & DDDDDDSUU (7+2 per 5~ms)                                                                                                                                                                                                                                                                                                                                                                                                                           \\
    gNB Rx antennas ($N_{\mathrm{rx}}$)          & 8, as a $2\times4$ half-$\lambda$ uniform planar array (UPA): 4 columns along azimuth, 2 rows along elevation                                                                                                                                                                                                                                                                                                                                      \\
    nrUE Tx antennas ($N_{\mathrm{tx}}$)         & 1 (single SRS port)                                                                                                                                                                                                                                                                                                                                                                     \\
    Antenna orientation                          & gNB Rx: azimuth baseline along $y$, boresight $+x$ (broadside to the corridor); elevation baseline vertical                                                                                                                                                                                                                                                                                   \\
    Antenna element pattern                      & Isotropic; both gNB Rx and nrUE Tx                                                                                                                                                                                                                                                                                                                                                                                                                 \\
    \midrule
    SRS occupied tones ($M_{\mathrm{SRS}}$)      & 624 (comb-2, every 2nd subcarrier over 106~PRBs; 1272 subcarriers total)                                                                                                                                                                                                                                                                                                                                                                           \\
    SRS periodicity ($T_{\mathrm{SRS}}$)         & \SI{10}{\milli\second}                                                                                                                                                                                                                                                                                                                                                                                                                             \\
    CPI length ($N_{\mathrm{occ}}$)              & 64~occasions                                                                                                                                                                                                                                                                                                                                                                                                                                       \\
    CPI duration ($T_{\mathrm{CPI}}$)            & \SI{640}{\milli\second}                                                                                                                                                                                                                                                                                                                                                                                                                            \\
    Range resolution ($\Delta R_{\mathrm{res}}$) & \SI{\MTrangeRes}{\metre} ($c/B_{\mathrm{SRS}}$, $B_{\mathrm{SRS}} \approx \SI{\MTbsrs}{\mega\hertz}$ SRS aperture, total bistatic path); zero-padded bin \SI{1.22}{\metre}                                                                                                                                                                                                                                                                                    \\
    Doppler resolution ($\Delta v$)              & \SI{0.14}{\metre\per\second} ($\lambda/T_{\mathrm{CPI}}$); \SI{0.07}{\metre\per\second} zero-padded bin                                                                                                                                                                                                                                                                                                                                            \\
    Wavelength ($\lambda$)                       & \SI{0.0903}{\metre}                                                                                                                                                                                                                                                                                                                                                                                                                                \\
    \midrule
    Clutter-deflation rank ($K_c$)               & 1 (held; a higher rank risks canceling the target, Section~\ref{sec:clutter})                                                                                                                                                                                                                                                                   \\
    CFAR type                                    & Order-statistic (OS-CFAR), $75^{\mathrm{th}}$-percentile training                                                                                                                                                                                                                                                                                                                                                                                  \\
    CFAR guard (range, Doppler)                  & 14, 5 cells (comb-LS range response)                                                                                                                                                                                                                                                                                                                                                                                                               \\
    CFAR training (range, Doppler)               & 28, 8 cells                                                                                                                                                                                                                                                                                                                                                                                                                                        \\
    CFAR $P_{\mathrm{FA}}$                       & $10^{-4}$                                                                                                                                                                                                                                                                                                                                                                                                                                          \\
    Single-peak non-maximum-suppression (NMS) range radius & 14 range bins (reduced from the single-target value of 18)                                                                                                                                                                                                                                                                                                        \\
    Raw CFAR crosser list                        & 512 (Section~\ref{sec:mt:rawcap})                                                                                                                                                                                                                                                                                                                  \\
    Max detections per CPI                       & 16                                                                                                                                                                                                                                                                                                                                                                                                                                                 \\
    \midrule
    EKF $\sigma_R$                               & \SI{6.0}{\metre} (re-derived, Section~\ref{sec:measmodel})                                                                                                                                                                                                                                                                                                                             \\
    EKF range bias correction                    & \SI{\MTrangeBias}{\metre}. \emph{Configured but not in force in the reported campaign}; every bistatic range figure carries the uncorrected offset (Section~\ref{sec:measmodel})                                                                                                                                                                                                                                                                                                                                                             \\
    EKF $\sigma_v$                               & \SI{0.89}{\metre\per\second} (measured range-rate residual)                                                                                                                                                                                                                                                                                                                                                                                        \\
    EKF $\sigma_{\theta_{\mathrm{az}}}$          & $\ang{6}$ (re-derived, Section~\ref{sec:measmodel})                                                                                                                                                                                                                                                                                      \\
    EKF $\sigma_a$ (process noise)               & \SI{0.5}{\metre\per\second^2}                                                                                                                                                                                                                                                                                                                                                                                                                      \\
    Mahalanobis gate ($\gamma$)                  & 20                                                                                                                                                                                                                                                                                                                                                                                                                                                 \\
    Association                                  & probabilistic data association, with birth inhibition and a $\ang{16}$ elevation birth separation (Section~\ref{sec:association})                                                                                                                                                                                                                                                                                                                  \\
    Coast limit                                  & 5~CPIs                                                                                                                                                                                                                                                                                                                                                                                                                                             \\
    Height prior $h_{\mathrm{tgt}}$ (2-D EKF)    & \SI{55}{\metre} --- a deployment parameter (expected altitude band), not inferred; removed by the 3-D upgrades (Sections~\ref{sec:upa}--\ref{sec:multistatic}) \\
    \bottomrule
  \end{tabular}
\end{table}

\begin{table}[!tbp]
  \centering
  \caption{Simulation Scene and Target}
  \label{tab:simtarget}
  \footnotesize
  \begin{tabular}{@{}p{0.40\linewidth}p{0.52\linewidth}@{}}
    \toprule
    \textbf{Parameter}                     & \textbf{Value}                                                                                                                                                                                                                                                                                                                                                                                                                     \\
    \midrule
    gNB / nrUE TX power (calibrated label) & 33 / 23~dBm (calibration-constant labels; \texttt{rfsimulator} transmits a fixed digital amplitude, not a power-control target)                                                                                                                                                            \\
    gNB / nrUE receiver noise figure       & 5 / 9~dB (thermal noise injected as calibrated AWGN at this NF; interference-over-thermal margin 0~dB)                                                                                                                                                                                                                        \\
    Uplink receive gain                    & \SI{\MTgainUl}{\decibel}, fixed at the controller's steady-state value; left adaptive it ramps between SRS bursts and over-drives the receiver (Section~\ref{sec:calib_validation}) \\
    \midrule
    Targets                                & \MTnTargets{} UAVs on parallel south--north transits at fixed $x$, diffuse spheres (scattering $0.9$); per-target values in Table~\ref{tab:mt:targets}. UAV-A ($x=27$~m, $z=\SI{55}{\metre}$, $v_y=\SI{2}{\metre\per\second}$) is the companion's single-target anchor~\cite{gurung_isac_testbed}, unchanged here                                                                                                                                                                                                                                                                                                                                                                                                                                                         \\
    gNB position                           & $(-75, 0, 10)$~m                                                                                                                                                                                                                                                                                                                                                                                                                   \\
    UE position (bistatic baseline)        & $(55, 0, 1.5)$~m, $L = \SI{130}{\metre}$                                                                                                                                                                                                                                                                                                                                                                                           \\
    Scene boundary ($x$, $y$)              & $[-100, 100]$~m, $[-60, 60]$~m                                                                                                                                                                                                                                                                                                                                                                                                     \\
    Channel model                          & Sionna RT (ray-traced; dry-ground floor, buildings removed for the flat-LOS benchmark)                                                                                                                                                                                                                                                                                                             \\
    \bottomrule
  \end{tabular}
\end{table}


Experiments are conducted using the parameters listed in
Tables~\ref{tab:params} and~\ref{tab:simtarget}. The gNB is placed at $(-75, 0, 10)$~m and the
nrUE at $(55, 0, 1.5)$~m, giving a bistatic baseline
$L = \SI{130}{\metre}$ along the main street corridor. A UAV
performs a horizontal transit at fixed $x = 27$~m and altitude
$z = \SI{55}{\metre}$ (above rooftop level, line-of-sight), flying
south-to-north along $+y$ at $\SI{2}{\metre\per\second}$ from
$y = -40$~m to $y = +42$~m. This trajectory carries the target through
the array broadside near $y = 0$, so a single run exercises the full
signature of interest: the bistatic range sweeps to a minimum and back,
the radial velocity reverses sign through the crossing, and the azimuth
sweeps through boresight. The UAV is modeled as a diffuse sphere
(radius $3$~m, scattering coefficient $0.9$), whose mean bistatic radar
cross-section measures \SI{+13.9}{\dBsm} on this single-target scene --- about
\SI{0.6}{\decibel} below its $\pi r^2$ optical limit of \SI{+14.5}{\dBsm}. (The
same mesh reads \SI{+13.4}{\dBsm} in the multi-target scene of
Section~\ref{sec:multitarget}: what the ray tracer returns is a bistatic aspect,
not a property of the mesh alone.) This is a deliberately strong
reference target; the detector characterization (Section~\ref{sec:detchar}, full
RCS ladder in the companion paper) scales it down to the declared band of
Section~\ref{sec:mt:geometry}
($-20$ to $-10$~dBsm) at which $P_{\mathrm{D}}$ is reported. Ground
truth positions are provided by the channel emulator's UAV
kinematic model and published to the key-value store for real-time comparison
with EKF estimates, and a run's sensor-timeline replay is additionally
rendered as an annotated trajectory video (ground truth vs. EKF estimate,
with the altitude panel) for qualitative inspection. The evaluation scene is the flat-LOS
benchmark of Table~\ref{tab:simtarget}; results use the Sionna
RT ray-traced channel model, providing ray-traced echo
coefficients (bistatic delay, Doppler shift, complex amplitude)
for the UAV at every simulation step.

\subsection{Calibration Validation and Single-Target Reference}
\label{sec:calib_validation}
\label{sec:baseline}

The platform's calibration and its single-target behavior are the companion
manuscript's subject~\cite{gurung_isac_testbed}; both are summarized here only far
enough to establish that the multi-target results sit on a physically calibrated
link and a characterized chain.

\emph{Link budget.} Three anchors computed from the constants the emulator uses
at runtime place a representative UAV echo well below the injected thermal floor
($\approx\SI{-93.2}{\dBm}$ at $\mathrm{NF}=\SI{5}{\decibel}$), recovered only
through the $\sim\SI{50}{\decibel}$ coherent processing gain of the CPI --- the
expected regime for a physically calibrated ISAC link. The full budget is in
Appendix~\ref{app:standalone}.

\emph{Uplink receive gain.} The receive gain is fixed by measurement rather than
inherited from the emulator's adaptive loop, which ratchets upward between the
\SI{\MTsrsGap}{\milli\second} SRS bursts and over-drives the receiver; pinning it
at \SI{\MTgainUl}{\decibel} moves the fraction of intervals clearing the CFAR
threshold from \MTposAuto\% to \MTposPinned\%. The value holds only for an uplink
carrying SRS alone.

\emph{Single-target reference.} Over $N=\CAMPflatN$ seeded runs the pipeline
detects in every observable CPI and tracks excess bistatic range to
\CAMPflatRangeR~m RMSE, range rate to \CAMPflatVelR~m/s and azimuth to
\ang{\CAMPflatAzR}. Two properties carry forward: the range figure sits at about
one \SI{8.0}{\metre} resolution cell (endfire-limited), and the transit is covered by \CAMPflatTrackIds{} identifiers rather
than one --- benign for a single target, but with several present a surplus
identifier becomes indistinguishable from a second target, the problem
Section~\ref{sec:association} addresses.

\emph{Role of this reference as context.} The run is single-target, on an
azimuth-only array, in a different scene, so no difference between it and the
multi-target results can be attributed to any one factor; the controlled baseline
is internal to the multi-target campaign (Table~\ref{tab:mt:basecmp}).


\textbf{Track fragmentation and coast-bridging.}
\label{sec:frag}
The identifier surplus is a \emph{publication} artifact as much as an estimation
one: a track whose next detection is late expires from the published set and
returns under a new identifier. A time-driven pass that forward-extrapolates each
confirmed track through such gaps therefore recovers continuity without touching
the estimator, and is the prerequisite the counter-UAS export of
Section~\ref{sec:c2} depends on.

The tracker is \emph{detection-driven}: it advances a track only on a detection
that is both delivered over the E2 service model for sensing (E2SM-SENS)
\emph{and} associated to
that track. Per-CPI detection is near-complete --- the target is detected on the
large majority of CPIs, with a median inter-detection gap of only
\SI{\CAMPflatDetGapMed}{\second} and a longest gap of \SI{\CAMPflatDetGapMax}{\second}
--- so the baseline's $100 - \CAMPflatContM \approx \SI{46}{\percent}$ continuity
shortfall is \emph{not} a detection problem but a track-management one. Azimuth
scatter from the four-element aperture spreads the per-CPI cross-range estimate
enough that detections intermittently fall outside the association gate; the
track then receives no update, its published record expires, and it must
re-confirm before it reappears --- the $\CAMPflatIdSwitch$ identifier switches per
run are the visible symptom. Each expire--reconfirm cycle
blanks the target picture even though the filter still holds a valid posterior.

We close these update gaps with a light \emph{time-driven} maintenance pass that,
once per second, forward-extrapolates each confirmed track's posterior along its
constant-velocity model and republishes it, flagged as a coasted (predicted, not
measured) estimate. Bridging is capped at \SI{\CAMPflatCoastMaxS}{\second}: beyond
that a pure prediction is no longer trustworthy, so the record is allowed to lapse
and the gap is reported honestly as a loss of track. The
pass reads filter state only --- it never advances the filter --- so the next
associated detection updates from the clean posterior and measurement accuracy is
unchanged.

With coast-bridging enabled, flat-LOS track continuity rises from
\CAMPflatCont\% to \CAMPflatContBridge\% ($N=\CAMPflatN$) \emph{on the wall clock}. On the
sensor's timeline the same five runs read \CAMPflatContSim\% and
\CAMPflatContBridgeSim\% --- a gain of \CAMPflatContSimDelta~percentage points
rather than \CAMPflatContWallDelta{} --- because the gap the pass exists to close
is opened by the emulator's dilation and not by the sensor
(Section~\ref{sec:c2_continuity}). Position accuracy, reported over
measured CPIs only (coasted predictions excluded), is unchanged within run-to-run
spread: range RMSE \CAMPflatRangeRmseBridge~m and cross-range
\CAMPflatXrangeRmseBridge~m. Bridging restores temporal continuity but not
single-identity tracking: the transit is still covered by \CAMPflatTrackIdsBridge{}
identifiers per run (\CAMPflatTrackIds{} without bridging), since republishing a
coasted track does not merge it with the fragment that later re-confirms ---
consolidating the fragments would need association-level work (e.g.\ JPDA). On either timeline the residual is track initiation and
re-confirmation latency --- the CPIs before a confirmed track first exists, which
a pass operating only on already-confirmed tracks cannot fill --- and not
detection dropouts, whose longest run (\SI{\CAMPflatDetGapMax}{\second}) sits well
inside the \SI{\CAMPflatCoastMaxS}{\second} bridging horizon. On the sensor
timeline that residual is about \SI{6}{\percent}, and it bounds what any
track-maintenance policy could add.

\subsection{Filter Consistency and Detector Characterization}
\label{sec:filter_consistency}
\label{sec:detchar}

A calibrated link budget says nothing about whether the tracker's assumed
uncertainty matches its actual error. The normalized innovation squared and
normalized estimation error squared test that, and both are defined and evaluated
for the single-target configuration in the
companion~\cite{gurung_isac_testbed}, where they establish that $\mathbf{R}$ must
be set from the outlier-inclusive residual spread. That result does not transfer,
and the reason is a contribution: three targets at differing
altitudes drive the filter through geometry the single-target case never visits,
and the statistics re-measured there (Section~\ref{sec:measmodel}) expose a
measurement-model error --- an azimuth convention valid only near broadside ---
that a single near-broadside target cannot reveal.

Detection uses OS-CFAR with the closed-form order-statistic multiplier and a
single scene-independent calibration offset, so the realized false-alarm rate
matches its nominal value ($\CAMPpfaAfter$ at a nominal $10^{-4}$); at the
calibrated operating point the pipeline detects the reference transit in
$\CAMPpdOpPd$ of observable CPIs at a median echo-cell SNR of $\CAMPpdOpSnr$~dB.
The full characterization --- $P_{\mathrm{D}}$ against echo-cell SNR, target RCS,
bistatic range and speed --- is the companion's, and its
$P_{\mathrm{D}}$-versus-echo-SNR curve is what the counter-UAS confidence model of
Section~\ref{sec:c2} calibrates against.

%
%
\section{Receive Configuration: Planar-Array Elevation}
\label{sec:upa}

The gNB's azimuth-only uniform linear array is replaced by a
$\UPArows\times\UPAcols$ ($\UPAelems$-element) uniform planar array at
$\lambda/2$ spacing, and the nrUE is reduced to a single SRS transmit port so the
angle estimate is formed entirely at the gNB receive aperture. A rectangular
array is separable: the inter-element phase ramp of~\eqref{eq:aoa} applies
independently along each baseline, giving azimuth from the horizontal ramp and
elevation from the vertical one at no additional estimator complexity, and
reducing exactly to the baseline azimuth estimator when the array is
one-dimensional. The tracker is extended to a three-dimensional
constant-velocity state with measurement
$[\,R_{\mathrm{bi}}, v_{\mathrm{bi}}, \theta_{\mathrm{az}},
\theta_{\mathrm{el}}\,]$, so the height prior $h_{\mathrm{tgt}}$ is removed and
altitude is estimated online; on the single-target benchmark this resolves
altitude to $\UPAcampHeightRmse \pm \UPAcampHeightSD$~m height
RMSE~\cite{gurung_isac_testbed}.

Two properties carry into the multi-target case, and both are limitations rather
than capabilities.

\emph{Elevation does not separate targets the detector merged.} Detection and
non-maximum suppression operate on the range--Doppler map alone, and the angle
estimator is invoked once per detected cell with a rank-one single-snapshot
covariance. Two targets occupying one range--Doppler cell therefore return one
detection carrying one blended bearing, in elevation exactly as in azimuth.
Multi-source angle estimation would require a covariance of rank greater than
one---spatial smoothing or subarray averaging---and is outside the scope of this
work. What elevation contributes is not resolution but \emph{association}: among
detections the detector has already resolved, it is the deciding discriminant in
\MTrendElOnly\% of intervals (Section~\ref{sec:association}).

\emph{Elevation accuracy is strongly geometry-dependent.} Per-CPI elevation error
varies by an order of magnitude along a single transit---
$\ang{\UPAliveEndfireEl}$ RMSE near endfire, where the vertical cut through the
target is most favorable, against $\ang{\UPAliveBroadsideEl}$ at broadside,
where a two-row aperture has least leverage on the vertical
angle~\cite{gurung_isac_testbed}. The meter-level altitude quoted above is therefore
not per-measurement accuracy but the product of a filter integrating many noisy
bearings across a transit. The consequence for the weakest target in the
multi-target scene is quantified in Section~\ref{sec:measmodel:elcorr}, where the
filter is made to decline to publish an altitude it cannot support rather than
publish a confident wrong one.

%
%
\section{Multistatic Geometry as an Alternative Route to Altitude}
\label{sec:multistatic}

Altitude can also be made observable without any vertical aperture, by adding a
second transmitter. A single gNB receiver served by two spatially separated nrUEs
observes the target over two distinct bistatic ellipsoids; where one pair
constrains the target to an iso-range surface, the intersection of two such
surfaces---together with the gNB-side azimuth---pins it in three dimensions. This
is attractive because it needs no change to the radio unit: a second commodity
UE, not a wider array. The companion manuscript~\cite{gurung_isac_testbed} derives the
fusion, reports the live campaign, and characterizes the geometry-dilution and
range-gate trade-off that governs it.

\section{Multi-Target Sensing}
\label{sec:multitarget}

This section establishes what the sensing front end can and cannot resolve with
\MTnTargets{} simultaneous UAVs on the \MTconfig{} configuration;
Section~\ref{sec:association} then evaluates the tracker on that input. \emph{Detectability},
not resolution, is the binding constraint on scenario design, and the detector's
capacity limit is a fixed-size internal buffer rather than anything about the
waveform.

\subsection{Scenario Design and the Resolution Yardstick}
\label{sec:mt:geometry}

The scenario places \MTnTargets{} UAVs of different size, altitude and velocity
in the bistatic footprint simultaneously (Table~\ref{tab:mt:targets}). The target
radar cross sections are set to a \emph{declared} band spanning
$\MTaRcs$ to $\MTcRcs$~dBsm by
rescaling the rendered channel to an absolute level. Untouched, the three spheres of
this scene measure $+13.4$, $+8.3$ and $+9.6$~dBsm --- large-aircraft scale.

\paragraph{Interpretation of the cross-section band}
It is a \emph{sensitivity scenario}: three echo levels a decade apart, imposed so
that the weakest target is demonstrably the one the detector loses first, and so
that a reader can locate this evaluation on an echo-strength axis. It is
\emph{not} a calibrated drone model. The
scattering primitive is a diffuse sphere, and the physical chain that would link
a sphere to an airframe --- aspect and polarization dependence at
\SI{3.32}{\giga\hertz}, body-versus-rotor decomposition, the spread of
measured bistatic cross sections across airframe classes --- is not established
here and cannot be established from a sphere. What the band \emph{does} support
is a relative statement: the ordering of the three targets, and the fact that
detection frequency is a joint function of cross section and geometry rather than
of cross section alone (Section~\ref{sec:mt:rawcap}). Restating any absolute
detection range for a named airframe would require measured bistatic UAV cross
sections at this frequency.

\begin{figure*}[!tbp]
  \centering
  \includegraphics[width=\textwidth]{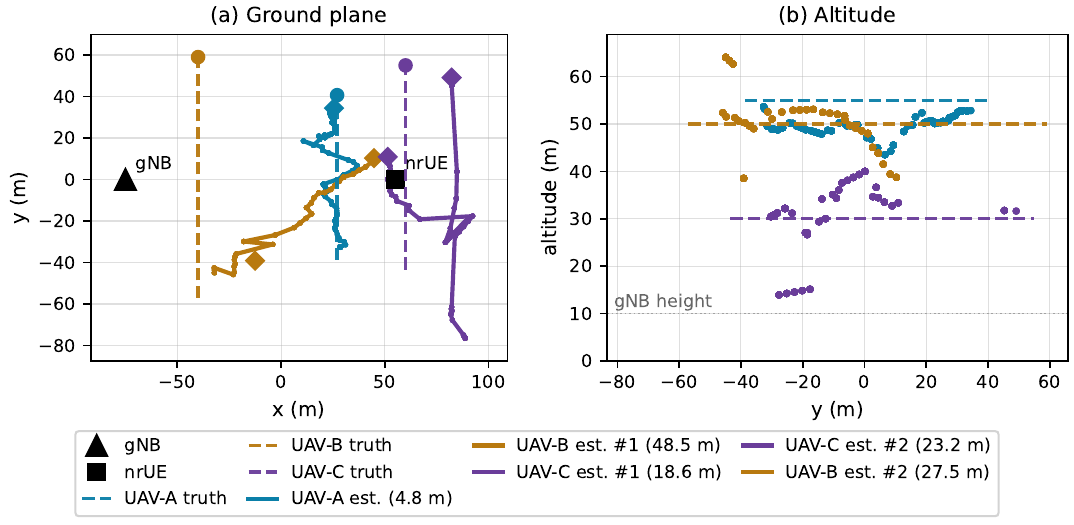}
  \caption{The \MTnTargets-target scene and its tracks, from the rendered trace. \emph{(a)} Plan view (ENU ground plane): gNB, nrUE, and three parallel south--north transits at fixed $x$. \emph{(b)} Altitude against $y$. Estimates (solid) are colored by assigned identity and shown against ray-traced truth (dashed); gaps are CPIs with no live track. One run (seed 1004), sensor-timeline replay (Section~\ref{sec:cadence}); campaign figures are in Table~\ref{tab:mt:tracking}.}
  \label{fig:track_overlay}
\end{figure*}

Those three numbers are also why \emph{mesh radius is not a usable cross-section
knob}. They are not monotonic in radius: the \SI{1.5}{\metre} sphere reads
\SI{1.3}{\decibel} \emph{higher} than the \SI{2}{\metre} one, because what the ray
tracer returns is the bistatic aspect of that mesh in that geometry rather than a
free-space $\pi r^2$. Shrinking a mesh therefore does not reliably shrink its
echo, and the absolute level has to be imposed afterwards by rescaling the
rendered channel --- which is what Table~\ref{tab:mt:targets} reports.
The ground truth of Fig.~\ref{fig:track_overlay} (dashed) shows the resulting
geometry.


\begin{table}[t]
\caption{Multi-target scenario. RCS is the equivalent bistatic cross section of
the rendered channel, set per target by rescaling. ``Notch'' is the fraction of
live CPIs spent inside the zero-Doppler blanking region.}
\label{tab:mt:targets}
\centering
\begin{tabular}{lrrrrr}
\toprule
Target & Alt. & RCS & Elev. & $\Delta R$ & Notch \\
 & (m) & (dBsm) & (\si{\degree}) & (m) & (\%) \\
\midrule
UAV-A & \MTaAlt & \MTaRcs & \MTaEl & \MTaRangeSpan & \MTaNotch \\
UAV-B & \MTbAlt & \MTbRcs & \MTbEl & \MTbRangeSpan & \MTbNotch \\
UAV-C & \MTcAlt & \MTcRcs & \MTcEl & \MTcRangeSpan & \MTcNotch \\
\bottomrule
\end{tabular}
\end{table}

\paragraph{The resolution used to judge separability}
Two targets are resolvable in range only if their bistatic path lengths differ by
more than the true resolution $c/B = \SI{\MTrangeRes}{\metre}$ at the occupied
SRS aperture $B=\SI{\MTbsrs}{\mega\hertz}$. This must not be confused with the
\emph{zero-padded interpolation bin} of the range FFT, which is roughly
\SI{1.22}{\metre} and is a sampling interval, not a resolution.
The same care applies on the Doppler axis, where the resolution
$\lambda/T_{\mathrm{CPI}} = \SI{\MTdopRes}{\metre\per\second}$ and the FFT bin
spacing $2v_{\mathrm{Nyq}}/N_{\mathrm{dop}} =
\SI{\MTdopBin}{\metre\per\second}$ differ by the zero-padding factor of two.

\paragraph{Detectability as the binding constraint}
The CFAR stage blanks a guard band of Doppler \emph{bins} around zero to suppress
the static clutter ridge, giving a notch of
$\pm\SI{\MTnotch}{\metre\per\second}$. A target whose bistatic range rate lies
inside that notch is not merely hard to separate from its neighbors---it is not
reported at all. This dominated the first scenario we designed: it scored well on
every resolvability screen and nevertheless yielded only two
of three targets in a live run, because the slowest UAV spent \SI{38}{\percent} of
CPIs inside the notch. Scenario design must therefore clear the notch first and
optimize separability second; the geometry in Table~\ref{tab:mt:targets} holds
every target at or below \MTaNotch\% notch occupancy, worst for UAV-A.

\paragraph{Analytic design, validated against ray tracing}
Because a full render is expensive and the constraints above are geometric, the
scenario is designed analytically---bistatic range, range rate, azimuth and
elevation evaluated in closed form against the resolution cells, the range gates
and the notch---and only then rendered. On the geometry finally adopted, the
analytic prediction and the ray-traced trace agree to within one point
(\MTpredResolvable{}\% vs.\ \MTrendResolvable{}\% of CPIs with all pairs
resolvable; \MTpredElOnly{}\% vs.\ \MTrendElOnly{}\% where elevation is the only
discriminant), so the analytic model can be trusted for scenario design without
rendering.

\subsection{Role and Limits of the Elevation Channel}
\label{sec:mt:elevation}

The vertical aperture is not a third axis for separating targets, for the
architectural reason given in Section~\ref{sec:upa}: two targets in one
range--Doppler cell return one detection carrying one blended bearing, in
elevation exactly as in azimuth. Its contribution is narrower and is an
association one --- among detections the range--Doppler stage has \emph{already}
resolved, it decides which measurement belongs to which track, and on the adopted
geometry it is the deciding discriminant in \MTrendElOnly{}\% of CPIs, where the
targets are separated in range or Doppler but too close in azimuth for the
association gate alone.

\subsection{Detector Capacity: A Single-Target Buffer}
\label{sec:mt:rawcap}

With the scenario clearing the notch, the detector still reported at most
\MTdetsBefore{} targets per CPI. The cause is neither the waveform nor the
non-maximum suppression stage, but a fixed-size internal buffer: the list of raw
CFAR threshold crossings passed to the suppression stage held \MTrawCapOld{}
entries, and it ran full in \MTrawSatBefore{} CPIs. When the list is full the
implementation retains the strongest crossings---correct policy---but with three
targets of deliberately different size, the smallest is outranked by clutter and
direct-path range sidelobes and never enters the list, so the suppression stage
never sees it. Non-maximum suppression was excluded as the cause directly: it
merges a pair only when they fall inside both the range and Doppler windows, and
every pair in this geometry is separated by more than twice the Doppler window.

\emph{Experimental fix: an explicit target-count parameter}
Raising the constant would have removed this instance and left the class intact.
For the experiments reported here the detector takes the scene's target count
$N_T$ as a declared parameter and derives both capacities from it: the raw
crosser list and the cap
on reported detections per CPI. At $N_T=\MTnTargets{}$ these are
\MTrawCapDerived{} and \MTdetsCapDerived{} entries. $N_T$ is \emph{declared, not
estimated} --- the detector does not infer how many targets are present.

This is an instrumentation fix for a controlled experiment rather than a deployable detector design, and it does not generalize. A counter-UAS sensor does not know how many targets are in its
footprint; that is the question it exists to answer. The number is
time-varying, and an under-declaration is precisely the case that matters ---
the extra target is the one that is missed, silently, exactly as in the defect
this replaced. Declaring $N_T$ is defensible here only because the scene is
known by construction and because it lets the capacity variable be \emph{held
fixed} while contention and identity are measured; it would be indefensible in
the field.

\subsection{Real-Time Feasibility}
\label{sec:mt:realtime}

Each additional target adds one channel tap to the emulator's per-batch hot path,
so three UAVs plus the static taps require \MTtapsPerUav{} taps against a fixed
IQ deadline, and the planar array doubles the per-tap cost relative to the
azimuth-only array (measured ratio \MTperTapRatio{}, matching the arithmetic
ratio of the two matrix shapes). The resulting load is \MTloadVsStable$\times$
that of the validated single-target planar-array configuration. Since a
previously attempted configuration at roughly twice that load collapsed---the
emulator starving the UE into continuous random access---this margin required
verification. A live run confirms the configuration is
stable: \MTrachHealthy{} random-access retries and \MTrntiHealthy{} distinct
radio identities, against \MTrachStarved{} and \MTrntiStarved{} in the collapsed
case. Multi-target sensing on the planar array is therefore real-time feasible on
the same commodity host.

\subsection{Altitude Observability and Its Geometric Limits}
\label{sec:mt:altitude}

The planar array makes altitude observable in principle. It does not make it
observable everywhere, and the three-target geometry spans the boundary.

Altitude is recovered from the elevation angle as
$z = z_{\mathrm{gNB}} + r_{\mathrm{ground}}\tan(\mathrm{el})$, a relation whose
linearization is valid only while the elevation is large compared with its own
error. With a measured elevation standard deviation of \MTsigmaElMeas$^\circ$, the
three targets sit at elevation-to-error ratios of \MTaElRatio, \MTbElRatio{} and
\MTcElRatio. Restricting attention to frames in which the track is horizontally
well attached to its target---so that association error cannot be blamed---and
asking what fraction of altitude estimates fall inside their own stated
one-sigma interval, the two elevated targets are usable and the grazing target is
not: its estimate degrades as attachment improves, which is the signature of a
geometric rather than an associative limitation.

We therefore publish no altitude below twice the elevation error, which suppresses
\MTaZsupp\%, \MTbZsupp\% and \MTcZsupp\% of frames for the three targets and
leaves the grazing target's remaining estimates calibrated rather than confidently
wrong. Table~\ref{tab:mt:altitude} reports the outcome. The test is applied to the
\emph{measured} elevation rather than the track's own state, since the latter is a
function of the altitude being judged and a track whose height has drifted upward
would otherwise pass its own test.

\begin{table}[t]
\caption{Altitude on the sensor's timeline, per target. ``Within $1\sigma$'' is
68\% for a calibrated estimate; above is conservative, below is over-confident.
No target is over-confident.}
\label{tab:mt:altitude}
\centering
\begin{tabular}{lrrrrr}
\toprule
Target & el ($^\circ$) & Bias (m) & SD (m) & Within $1\sigma$ & Suppressed \\
\midrule
UAV-A & \MTaEl & \MTaZbias & \MTaZsd & \MTaZin\% & \MTaZsupp\% \\
UAV-B & \MTbEl & \MTbZbias & \MTbZsd & \MTbZin\% & \MTbZsupp\% \\
UAV-C & \MTcEl & \MTcZbias & \MTcZsd & \MTcZin\% & \MTcZsupp\% \\
\bottomrule
\end{tabular}
\end{table}

Only UAV-A supports an altitude claim without qualification. UAV-B's median error
is comparable but its spread is dominated by a heavy tail traced to duplicate
identity (Section~\ref{sec:cadence:residual}), and UAV-C's altitude is refused for
two thirds of its transit by the observability test above.

\section{Multi-Target Association and Track Identity}
\label{sec:association}

Section~\ref{sec:multitarget} establishes the detector's multi-target output:
several detections per CPI, of which a measurable fraction correspond to no
target. This section evaluates what the tracking xApp makes of that input, and
sets out the measurement discipline the evaluation requires.

\emph{Terms.} Several quantities are easily conflated under the word
``association'', and the results below depend on separating them.
Table~\ref{tab:mt:glossary} defines every metric used from here on, in the order
the chain applies them, together with what each is scored over --- the
distinction that most often causes two of them to be read as the same number.

\begin{table*}[t]
  \caption{Metrics used in Sections~\ref{sec:association}--\ref{sec:c2}. The
  first block is the failure chain, in the order it binds; the second is the
  set-level and identity measures. ``Scored over'' is what the denominator is,
  and is why availability, coverage and served fraction are not
  interchangeable.}
  \label{tab:mt:glossary}
  \footnotesize
  \setlength{\tabcolsep}{4pt}
  \begin{tabular}{@{}llp{0.44\linewidth}l@{}}
    \toprule
    Term & Answers & Definition as used here & Scored over \\
    \midrule
    \multicolumn{4}{@{}l}{\emph{The failure chain, in the order it binds}}\\
    \textbf{Sensitivity}       & Is it detectable?  & Whether the detector reports a target at all when it is alone in its resolution cell. & CPIs the target is present \\
    \textbf{Contention}        & Is the detection shared? & Whether two targets' nearest detection is the \emph{same} detection. Reported as the contested-cell fraction. & CPIs with $\ge 2$ targets \\
    \textbf{Availability}      & Does it get its own? & Whether a target has a detection of its own once contention is resolved by a one-to-one assignment. & CPIs the target is present \\
    \textbf{Admission}         & Does a track exist? & Whether a live track is created and kept for an available detection. & available detections \\
    \textbf{Association quality} & Is it the right one? & How well the tracker, given admitted tracks, assigns detections and preserves identity. & admitted tracks \\
    \midrule
    \multicolumn{4}{@{}l}{\emph{Set-level, accuracy and identity measures}}\\
    \textbf{Coverage}          & \multicolumn{2}{p{0.60\linewidth}}{Fraction of ground-truth samples carrying a matched estimate inside the \SI{20}{\metre} scoring gate. \emph{Bounded by that gate}; rises when a change publishes estimates on more of the scene.} & GT samples \\
    \textbf{Served fraction}   & \multicolumn{2}{p{0.60\linewidth}}{Of the CPIs in which a target had its own gating detection, the fraction in which it also held a track. Isolates admission from availability.} & detection-bearing CPIs \\
    \textbf{GOSPA}             & \multicolumn{2}{p{0.60\linewidth}}{Set distance folding localization error, missed and false targets into one quantity, decomposed into those three terms.} & CPIs \\
    \textbf{IDF1}              & \multicolumn{2}{p{0.60\linewidth}}{Identity F1 score: harmonic mean of identity precision and recall over the best global identity assignment.} & whole run \\
    \textbf{Identifiers / target} & \multicolumn{2}{p{0.60\linewidth}}{Distinct track identifiers a single target is carried by over its transit. Ideal is~1.} & per target, per run \\
    \textbf{Identity switches} & \multicolumn{2}{p{0.60\linewidth}}{Times an identifier moves from one target to another, or a target's identifier is re-minted.} & per run \\
    \textbf{Track continuity}  & \multicolumn{2}{p{0.60\linewidth}}{Fraction of the transit on which \emph{some} track exists for a target. \textbf{Inflated by fragmentation} --- see Section~\ref{sec:assoc:metrics}.} & CPIs the target is present \\
    \textbf{Primary-identity RMSE} & \multicolumn{2}{p{0.60\linewidth}}{Position error of the \emph{single} identity a consumer would be handed, over every frame in which it exists. Ungated, so unbounded.} & frames of one identity \\
    \bottomrule
  \end{tabular}
\end{table*}

The chain runs in that order. Sensitivity is adequate: no target is starved of raw
detections (Section~\ref{sec:assoc:availability}). Contention is nevertheless
high, and because contention consumes detections it is what \emph{reduces}
availability rather than sensitivity doing so. The loss remaining after
availability is accounted for is admission, not association quality
(Sections~\ref{sec:assoc:birth} and~\ref{sec:cadence:residual}). Association
quality itself the evidence exonerates: a geometry change that costs every target
availability leaves IDF1 and the set-level distance unmoved while
degrading the identifier count and switch rate
(Section~\ref{sec:assoc:heading})---the tracker keeps assigning the right
detection and keeps losing the identifier it does so under, which are separable
failures and are reported separately throughout.

\subsection{Association Under Multiple Targets}
\label{sec:assoc:pda}

The tracker maintains one extended Kalman filter per target and associates
detections probabilistically: for each track, all detections inside its validation
gate contribute to the update, weighted by association probabilities, rather than
a single winner being chosen. Two mechanisms carried over from the single-target
work matter more with several targets. Birth inhibition prevents a detection
already inside a live track's gate from spawning a competing identifier---the
dominant source of fragmentation---and an echo-strength fingerprint discourages
identifier exchange when two targets cross in measurement space, which the
heterogeneous cross sections of Table~\ref{tab:mt:targets} make discriminative.

The validation gate is deliberately not widened for the multi-target case:
widening buys fewer identifiers but admits angle-offset returns into the
association mixture, and in the single-target study that trade degraded
cross-range accuracy and filter consistency severely. The operating point is held
at the value established there.

\subsection{Measuring Identity}
\label{sec:assoc:metrics}

Three metrics carried by this testbed are inflated by fragmentation, because each
pools, per CPI, the estimate closest to the target across \emph{all} of that
target's fragments: track continuity, position RMSE, and filter consistency over
coasted records. That is a best-of-many statistic in which the number of
candidates is the quantity an
identity fix reduces, so a genuine improvement in identity presents as a
regression in all three. We therefore report, alongside the pooled figures, the
accuracy of the \emph{single} identity a downstream consumer would be handed, and
gate conclusions on identity-aware measures---identifier count, identity switches,
IDF1---rather than on continuity.

Two distinct radii appear throughout and are not interchangeable; the difference
decides whether a figure may be compared with an external requirement. The
\SI{100}{\metre} \emph{assignment} radius decides which track is attributed to
which target, and is deliberately permissive so that a drifting but real track is
credited to its target rather than discarded. The \SI{20}{\metre} \emph{scoring}
gate decides which matched pairs enter an average.

The consequence is the opposite of the intuitive one. \emph{Scoring at the wider
assignment radius is the conservative choice}: it admits the poor matches, so the
error is unbounded and can be read against an external requirement directly.
\emph{The tighter scoring gate is the flattering one}: an estimate
\SI{300}{\metre} from its target does not raise a gated mean, it leaves the
population and lowers coverage instead, so the statistic is bounded above by the
gate by construction. A gated mean therefore \emph{rises} when a change publishes
estimates on more of the scene, and that rise is selection, not degradation. Every
gated figure below is labeled as such, and no gated figure is compared with a
requirement.

A second artifact must be neutralized before any identifier count is meaningful.
The tracker historically allocated identifiers from a small fixed pool, which
censors the count at the pool size; with three targets the pool is saturated by
construction and fragmentation becomes unobservable. All results here use a
monotonic allocator. The pool also capped the number of \emph{concurrent} tracks,
so removing it admits genuine false tracks that were previously suppressed and the
false-track term of the set metric rises accordingly---the censoring being
removed, not a regression introduced by multiple targets.

\subsection{Set-Level Performance}
\label{sec:assoc:results}

\paragraph{Interpretation and scope of the statistical claims}
\label{sec:assoc:stats}
Each arm is \MTcampN{} noise seeds of \emph{one} deterministic
trajectory on \emph{one} geometry with \emph{one} gNB siting, and the only
geometric variation anywhere in the paper is a single alternative heading
(Section~\ref{sec:assoc:heading}). A $p$-value here therefore answers one
question --- whether an effect survives the noise realizations of this scenario
--- and cannot address how often it would appear across trajectories, altitudes,
speeds, target counts or sitings, which these samples do not sample. We
consequently adopt three rules. \emph{(i)} We lead with effect sizes and
dispersion; every quantity is quoted as mean~$\pm$~standard deviation across
seeds, and where an interval carries an argument it is given. \emph{(ii)} We
make no multiplicity correction and do not present these as controlled
family-wise error rates: dozens of metrics are examined across the paper, so an
isolated $|t|$ near the threshold should be read as descriptive, and we rest
conclusions only on effects that are large, mechanistically explained, and
consistent across the targets or arms they should be consistent across.
\emph{(iii)} Where the sample cannot support a direction we decline the claim
rather than pick one. The consequence is that the \emph{mechanisms} reported ---
wall-clock dilation changing a metric rather than delaying it, contention binding
before sensitivity, a gated mean rising when coverage rises, elevation deciding
association --- are argued from their causes and are what we expect to transfer,
while the \emph{magnitudes} are properties of this scenario. Read every
``separates'' below as ``separates on the noise seeds of this scene''.

Per-track RMSE alone cannot describe a multi-target estimate, being blind to
missed and spurious targets. We therefore report the GOSPA distance, which folds
localization error, missed
targets and false targets into one quantity and---unlike the optimal sub-pattern
assignment (OSPA) distance it generalizes---decomposes cleanly
into those three components without diluting cardinality error as the target count
grows.

Over $N=\MTcampN$ seeded transits of the \MTnTargets-target scenario, the tracker
attains GOSPA \MTgospa{} (localization \MTgospaLoc, missed \MTgospaMissed, false
\MTgospaFalse), with \MTidentifiers{} identifiers per target, \MTidsw{} identity
switches per transit and IDF1 \MTidfOne. Per-target position accuracy is
summarized in Table~\ref{tab:mt:tracking}. Detection probability varies across the
three targets---\MTaPd, \MTbPd{} and \MTcPd{} for UAV-A, UAV-B and UAV-C
respectively---reflecting the combination of cross section and notch occupancy
discussed in Section~\ref{sec:mt:rawcap} rather than cross section alone.

\paragraph{Definition of the control and three related comparisons}
A multi-target elevation claim needs a control that differs from it in elevation
alone. The single-target, azimuth-only campaign of the companion
paper~\cite{gurung_isac_testbed} is not that control: it differs in target count,
in array geometry and in scene at once, so no difference between it and the
numbers above can be attributed to the vertical aperture. It is context for this
work --- we call it the \emph{single-target reference} throughout, and it is
context rather than a control. Three further comparisons appear in this paper and
none is interchangeable with it: the \emph{companion platform validation} (the
cited single-target campaign that establishes the sensing chain), the
\emph{2-D height-prior control} defined immediately below, and the
\emph{timeline replay control} of Section~\ref{sec:cadence}, which re-scores one
campaign against itself. Only the second is the control for the results in this
section. It is internal---the same
\MTcampN{} runs, the same detections and the same tracker configuration, with the
filter's height prior replaced by an estimated altitude as the only variable.
Table~\ref{tab:mt:basecmp} reports it. Estimating altitude rather than assuming it
raises IDF1 from \MTbaseIdfOne{} to \MTidfOne{} and lowers horizontal RMSE from
\MTbaseRmse~m to \MTrmse~m ($p=\MTpIdf$ and $p=\MTpRmse$ on the paired seeds), and
raises coverage from \MTbaseCov\% to \MTcovSim\% ($p=\MTpCov$). What it does
\emph{not} improve is identity, and that is set out below.

\paragraph{Conditioning of the horizontal-RMSE row}
That RMSE is an average over \emph{matched} pairs, and is therefore bounded by the
\SI{20}{\metre} scoring gate in the way set out above. The two arms are not
averaged over the same samples --- the three-dimensional filter is matched on a
larger fraction of the scene, which is itself one of the results --- so the row
should not be read as an accuracy comparison in isolation. This is why the
set-level metric of Section~\ref{sec:assoc:results} is GOSPA, which prices a
missed target at a fixed cost and is immune to the same mechanism.

The conditioning is measurable. Rescoring the same runs and detections with the
match gate removed, and changing nothing else, the pooled figure rises from
\MTrmseGated~m to \MTrmseUngated~m for the three-dimensional filter and from
\MTbaseRmseGated~m to \MTbaseRmseUngated~m for the height prior, while the
fraction of ground-truth samples carrying any estimate rises from \MTcovGated\% to
\MTcovUngated\% and from \MTbaseCovGated\% to \MTbaseCovUngated\% respectively.
\emph{The gated advantage does not survive that change}: paired over the
$N=\MTcampN$ shared seeds it is \MTrmseDeltaGated~m at the gate
($p=\MTrmsePGated$, exact sign-flip) and \MTrmseDeltaUngated~m without it
($p=\MTrmsePUngated$, not separated). The reversal is selection, not regression:
the three-dimensional filter publishes an estimate on eight percentage points more
of the scene, and the samples it adds are the difficult ones the two-dimensional
filter never attempted, so a mean over published samples penalizes the arm that
attempts more. We report the gated figure because every metric beside it is scored
at the same gate, and state its conditioning rather than let the row carry an
unconditional reading.

\paragraph{Composition and poolability of the $N=\MTcampN$ campaign}
Each campaign in this paper is two batches of \MTcampBatch{} seeds rather than one
campaign of \MTcampN, and the batches were not produced by the same repository
revision. On the baseline scene one tracker calibration constant gained
configuration passthrough between them, so the second batch's \emph{live} xApp
ran with a \SI{7.5}{\metre} different range-bias correction. We disclose it
because a top-up described as adding seeds in fact added a configuration change.

It does not invalidate the pool, for a specific reason: every metric quoted here
is re-derived from the \emph{gNB logs} under one environment rather
than read from the xApp's output, and the gNB, channel-emulator and RIC image
digests are byte-identical across the two batches --- only the discarded xApp
output differs. A mechanism argument is not a measurement, so each split was also
tested directly. Permuting the batch labels over all \MTpermSplits{} distinct
five-and-five relabellings, of \MTpermTested{} comparable metrics the baseline's
true split flags \MTpermFlagged{} at $|t|>2.31$, where a random split flags a
median of \MTpermMedian{} and as many as \MTpermMax{} (permutation
$p=\MTpermP$). The alternative-geometry campaign was tested the same way and is
cleaner still: \MTpermFlaggedDiag{} of \MTpermTestedDiag{} against a random
median of \MTpermMedianDiag{} (permutation $p=\MTpermPDiag$). By that test each
revision boundary is an ordinary partition of its \MTcampN{} runs. The
one exception is identity switches on the baseline (\MTidswBatchOld{} against
\MTidswBatchNew, $p=\MTidswBatchP$ --- about what \MTpermTested{} tests produce
on their own, which is why we do not treat it as a finding); consistent with
that, no identity claim in this paper now rests on a difference between the arms,
because none survives the doubling.

The test is a check on the replay path, and it is only that. A metric read
directly from the xApp's recorded output has not passed through the thing the
permutation tested, and one such quantity --- the latency to hold a third
concurrent track --- does split perfectly on the revision boundary. No figure in
this paper is taken from that path.

The accuracy claim that does not depend on the gate is primary-identity RMSE,
accumulated over every frame in which the assigned identity exists and not gated
at all. It favors the three-dimensional filter at both gates and reverses at
neither.

The mechanism is visible per target rather than only in the aggregate. The
scenario places its three UAVs at \MTaAlt, \MTbAlt{} and \MTcAlt~m while the
two-dimensional filter carries one height prior, so the lowest target is the one
the prior misdescribes most---and it is the one that gains most, its served
fraction rising from \MTbaseCservedGtGated\% to \MTcservedGtGated\%
(Section~\ref{sec:assoc:birth}, where the gate these are scored at is stated).

\begin{table*}[!tbp]
\caption{The \emph{2-D height-prior control}: same \MTcampN{} runs, detections and tracker; only the filter's height prior is replaced by an estimated altitude. $p$ is the exact paired sign-flip test, uncorrected for multiplicity (Section~\ref{sec:assoc:results}). Rows marked $\dagger$ are matched-pair averages bounded by the \SI{20}{\metre} gate; that horizontal-RMSE advantage does not survive gate removal (text). Neither identity row separates.
}
\label{tab:mt:basecmp}
\centering
\begin{tabular}{@{}lrrrl@{}}
\toprule
 & 2-D, height prior & 3-D, elevation & $p$ & \\
\midrule
Coverage (\%)                 & \MTbaseCov            & \MTcovPooled      & \MTpCov         & \\
IDF1                          & \MTbaseIdfOne         & \MTidfOne         & \MTpIdf         & \\
Horizontal RMSE (m)$^\dagger$ & \MTbaseRmse           & \MTrmse           & \MTpRmse        & at the \SI{20}{\metre} gate only \\
UAV-C served (\%)$^\dagger$   & \MTbaseCservedGtGated & \MTcservedGtGated & \MTpCserved     & \\
Primary-identity RMSE (m)     & \MTbasePrimaryRmse    & \MTprimaryRmse    & \MTpPrimary     & \\
Identifiers / target          & \MTbaseIdentifiers    & \MTidentifiers    & \MTpIdentifiers & does not separate \\
Identity switches             & \MTbaseIdsw           & \MTidsw           & \MTpIdsw        & does not separate; sign reversed \\
\bottomrule
\end{tabular}
\end{table*}

\paragraph{Effect of the larger sample on the two identity rows}
At $N=\MTcampBatch$ the two identity rows were the strongest results in it:
identifiers per target separated at $|t|=4.0$ and identity switches at $|t|=3.2$,
both favoring the three-dimensional filter, and an earlier version of this paper
reported them as an identity improvement. At $N=\MTcampN$ neither separates, and
identity switches point the other way (\MTbaseIdsw{} for the height prior against
\MTidsw). We withdraw that claim. It is not a close call about a threshold: the
effect the smaller sample showed is not present in the larger one, and reporting
the smaller sample because it was more favorable would be the wrong way round.
Coverage and primary-identity RMSE moved the other way --- neither separated
cleanly at $N=\MTcampBatch$ and both do now --- which is what doubling a sample
is supposed to do in both directions.

Every figure above is scored on the \emph{sensor's} timeline --- the
\SI{\MTcpiPeriod}{\second} CPI period at which the scene was actually sampled ---
not on the host's wall clock, for the reason Section~\ref{sec:cadence} gives: two
constants in the evaluation path carry wall-clock units, so wall-clock scoring
imposes a coverage ceiling of \MTceilingWall\% under which ``mostly tracked'' is
unreachable by construction. The same detections driving the same tracker score
coverage \MTcovWall\% against \MTcovSim\%, and fragmentation \MTfmWall{} against
\MTfmSim{} per transit, purely from that choice.

The GOSPA decomposition inverts with it, which changes the diagnosis rather than
the presentation. On the wall clock the distance is dominated by the missed term
(\MTgospaMissedWall{} of \MTgospaLocWall{}/\MTgospaFalseWall{} localization and
false), inviting the reading that the tracker loses targets. On the sensor
timeline it is a three-way split (\MTgospaMissedSim{} missed, \MTgospaLocSim{}
localization, \MTgospaFalseSim{} false): most wall-clock polls had no live track
at all, so there was nothing to localize and nothing to report falsely. A
substantial part of the set-level error is therefore \emph{spurious tracks}, not
absent ones --- which points at the admission rule of
Section~\ref{sec:assoc:birth} rather than at detection.

\subsection{Availability and the Role of Contention}
\label{sec:assoc:availability}

Before attributing any deficit to the tracker, we establish what the detector
makes available per target. The natural measurement---how often some detection
falls inside a validation gate placed at a target's true state---overstates
availability, because two targets in one resolution cell both ``gate'' on the same
detection. Scored instead under a one-to-one assignment, so a detection can serve
at most one target, the three have their own gating detection in \MTaExcl\%,
\MTbExcl\% and \MTcExcl\% of CPIs. Availability does not order by
cross section: the \emph{largest} target is the least available, because it also
spends the largest fraction of the transit inside the zero-Doppler notch
(\MTaNotch\% against \MTbNotch\% and \MTcNotch\%). The weakest-cross-section
target is not detection-starved, and no target is available in fewer than about
two-thirds of intervals. Sensitivity is therefore not what limits this scenario.

The contested fraction of CPIs---those in which two targets' nearest detection is
the same one---is \MTcontestedNoEl\% in range, Doppler and azimuth alone, falling
to \MTcontestedEl\% when elevation joins the assignment. This is the quantitative
form of the planar array's contribution to \emph{association}, as distinct from
its contribution to localization: it is the discriminant in the cells where the
other three dimensions have already collapsed. The two sides are scored on the
same $N=\MTcontestedElN$ runs and their configurations differ in one setting
only---whether elevation enters the assignment---so the \emph{difference} is what
the contrast measures, and the elevation-enabled half is by construction the
unrestricted \MTcontested\% quoted elsewhere.

\emph{Frequency of elevation-decisive intervals over sampled rather than designed geometry.}
The figures above are measured on a scenario built to place three targets in a
sensible spread, and a possible objection is that the configuration was selected
for it. The frequency question is separable from the sufficiency question, and we
answer it by Monte Carlo over trajectories \emph{drawn} from a stated envelope
rather than designed: level flight, uniform heading, \SIrange{15}{120}{\metre}
above ground level (AGL), and the \SIrange{1.5}{3.5}{\metre\per\second} speed
window this testbed can
render, with the same cells and angular sigmas the scenario design tool uses. A
pair counts as contested when the detector resolved it into two detections --- a
precondition, since two targets sharing a cell yield one detection carrying one
blended bearing, against which a vertical aperture is powerless --- but azimuth
alone cannot associate them.

About \MTelContestedNat\% of jointly reportable intervals are contested in that
sense, and elevation is the deciding discriminant in \MTelBaseRate\% of them
(\MTelBaseRateLo--\MTelBaseRateHi\% over \MTelSeeds{} independent draws): not a
majority, and not the corner case the objection supposes. We quote the
unconditional rate because the conditional one --- the fraction of contested
intervals elevation rescues --- is far less stable across draws and cannot be
quoted responsibly without replication.

\subsection{Birth Admission Under Stacked Targets}
\label{sec:assoc:birth}

Birth inhibition most requires revisiting for multiple targets, because it was
tuned where its only job was to stop one target minting duplicate identifiers.
Both of its vetoes are blind to altitude---one operates on range, Doppler and
azimuth, the other on ground-plane separation---so two UAVs stacked over nearly
the same ground track are indistinguishable to it, and the second is suppressed
for the whole transit.

The three-dimensional filter addresses this at its source, and is sufficient on
its own. Under the height prior the lowest target is served by a live track in
only \MTbaseCservedGtGated\% of detection-bearing CPIs despite having its own
gating detection in \MTcExcl\% of them; estimating altitude raises that to
\MTcservedGtGated\%. The incumbent's predicted bistatic range is then right for a
target at its own height, so the veto no longer fires against a genuinely separate
UAV, and no special case is required. The gain is a factor of three, but it leaves
the lowest target served for only a quarter of the CPIs in which it was detected:
the birth veto is no longer the binding constraint on that target, and something
downstream of it now is.

\paragraph{Geometry as the binding constraint}
Neither remaining loss is an association failure. Both targets that fall short
have tracks; those tracks sit outside the \SI{20}{\metre} scoring gate, at a
total position error of \MTbTotalRmse{} and \MTcTotalRmse~m against
\MTaTotalRmse~m for the target that does not fall short --- which is also why the
permissive assignment radius concealed the losses, since an error of that size
still attributes a track to its target. The two fail different screens.

UAV-B fails in azimuth. It flies \SI{\MTbGndRange}{\metre} from the gNB, so its
transit sweeps $\pm\ang{\MTbAzMax}$ where the other two sweep about
$\pm\ang{\MTaAzMax}$, and at its excursions the effective aperture is halved
(\MTbAperture{} against \MTaAperture{} and \MTcAperture). The array was placed
where it is precisely to keep targets near broadside; this is the one target
placed where that does not hold, and its cross-range error follows.

UAV-C fails in elevation, and not for a fixable reason. Its elevation subtends
only \MTcElK$\sigma_{\mathrm{el}}$, against the $\MTelObsK\sigma$ an altitude
needs before the geometry supports one. That threshold is a \emph{design}
criterion here, and the distinction matters for reading the rows above: the
tracker implements it as an optional gate that withholds an altitude below
$\MTelObsK\sigma_{\mathrm{el}}$, and \textbf{that gate was not enabled in these
campaigns} --- every altitude they produced was published. This is why UAV-C has
a measurable altitude row at all (\MTcAltRmse~m) despite sitting far below the
floor, and the row should be read as what the estimator does when it is
\emph{not} allowed to decline: at \MTcElK$\sigma$ the answer is that it should
have. The limit is structural rather than incidental, and can be set out as two
competing constraints on one free variable, the target's ground range $r$ from
the array.

\emph{Elevation observability wants $r$ small.} The subtended elevation is
$\theta_{\mathrm{el}} = \arctan\big((z - z_{\mathrm{gNB}})/r\big)$, so for a
fixed height it falls monotonically as the target moves out. Publishing an
altitude requires $\theta_{\mathrm{el}} > \MTelObsK\sigma_{\mathrm{el}}$,
which is an \emph{upper} bound on $r$:
\begin{equation}
  r \;<\; \frac{z - z_{\mathrm{gNB}}}
                {\tan\!\big(\MTelObsK\,\sigma_{\mathrm{el}}\big)}.
  \label{eq:elobs}
\end{equation}

\emph{Broadside azimuth wants $r$ large.} A transit of half-extent $y_{\max}$ at
ground range $r$ sweeps azimuth to $\pm\arctan(y_{\max}/r)$, and the
interferometric estimator degrades toward endfire as the effective aperture
$\cos\theta_{\mathrm{az}}$ shrinks. Holding the excursion within
$\theta_{\mathrm{az}}^{\max}$ is a \emph{lower} bound,
$r > y_{\max}/\tan\theta_{\mathrm{az}}^{\max}$.

The two are simultaneously satisfiable only when the upper bound exceeds the
lower, which rearranges to a condition on height alone:
\begin{equation}
  z \;>\; z_{\mathrm{gNB}}
      + y_{\max}\,
        \frac{\tan\!\big(\MTelObsK\,\sigma_{\mathrm{el}}\big)}
             {\tan\theta_{\mathrm{az}}^{\max}} .
  \label{eq:elaz}
\end{equation}
A target below that height has no ground range at which both hold: pull it in for
elevation and it swings off broadside; push it out for azimuth and its elevation
falls under the observability floor. UAV-C at \MTcAlt~m is below it, UAV-A at
\MTaAlt~m is above it, and no tracker change moves the boundary --- only the
geometry does. The low, fast-crossing target is precisely the one the
vertical aperture exists to serve. This is the measured limit of a single planar
array rather than a deficiency of the tracker, and it is the boundary a second
bistatic pair is intended to move (Section~\ref{sec:multistatic}).

\paragraph{The elevation residual as scatter rather than a calibratable offset}
An obvious objection to the preceding paragraph is that the array carries a known
elevation bias --- earlier characterization of this pipeline measured
\SIrange{\MTelPriorArmLo}{\MTelPriorArmHi}{\degree} on one build and
\SIrange{\MTelPriorXHi}{\MTelPriorXLo}{\degree} on another, on the same trace and
configuration --- and that subtracting it would recover the altitudes. That
characterization was made on a single target at one elevation, which cannot
distinguish a constant from a slope. Three targets at three altitudes, observed
in the same CPI by the same build, can. Matching detections to truth on bistatic
range-rate alone --- never on elevation, which would make the residual circular,
nor on azimuth, which compresses off-boresight and would preferentially reject
the highest target --- and restricting to the CPIs in which no other target's
\emph{true} range-rate lies inside the association gate, the residual is
\SI{\MTelBiasC}{\degree} at \SI{\MTelTrueC}{\degree} of true elevation,
\SI{\MTelBiasA}{\degree} at \SI{\MTelTrueA}{\degree} and
\SI{\MTelBiasB}{\degree} at \SI{\MTelTrueB}{\degree}: three targets agreeing to
\SI{\MTelBiasSpread}{\degree}, each smaller than the offset the objection
proposes to remove. The dependence on the angle is
\MTelSlope{} degrees of residual per degree of elevation (95\,\% CI
\numrange{\MTelSlopeLo}{\MTelSlopeHi}, bootstrapped over runs), which across the
\SIrange{\MTelSpanLo}{\MTelSpanHi}{\degree} the scene spans is worth
\SI{\MTelSlopeSwing}{\degree} end to end --- resolvable, and smaller than the
\SI{\MTelRmseClean}{\degree} per-sample scatter it sits inside. Fitting a
correction on nine runs and scoring it on the tenth confirms the consequence:
\SI{\MTelRmseClean}{\degree} uncorrected, \SI{\MTelCalConst}{\degree} with a
constant, \SI{\MTelCalLin}{\degree} with a term linear in the angle,
\SI{\MTelCalTgt}{\degree} with a per-target constant --- at most
\SI{\MTelCalBestPct}{\percent}. The residual is scatter, and scatter does not
subtract.

What \emph{does} dominate the elevation error is the scene rather than the
sensor. In the \SI{\MTelAmbPct}{\percent} of samples where another target sits
inside the association gate, UAV-C reads \SI{\MTelAmbBiasC}{\degree} with
\SI{\MTelAmbSdC}{\degree} of scatter, and the held-out error rises from
\SI{\MTelRmseClean}{\degree} to \SI{\MTelRmseAmb}{\degree}. In that stratum a
constant correction appears to buy a fifth of the error and a linear one more
than a third. That apparent gain should not be interpreted as a correction: it is
a fit to this scene's target geometry, and
it would invert the moment the weak target changed altitude. The offset is a
property of neither the angle nor the build in any form that transfers, and we
therefore report altitudes uncorrected.

\paragraph{Accuracy as a siting-dependent result}
Combining the two error components gives a total position error of
\SI{\MTaPosRmse}{\metre}, \SI{\MTbPosRmse}{\metre} and \SI{\MTcPosRmse}{\metre}
for UAV-A, UAV-B and UAV-C, unbounded by the \SI{20}{\metre} gate --- unlike the
pooled figure of Section~\ref{sec:assoc:results} --- so they can be read against
an external requirement directly. Each is scored over the intervals in which the
target's own track was the nearest estimate to it, for the reason given below.

The \MTposSpread$\times$ spread across the three is the informative part. All
three are scored on the \emph{same} runs, from the same detections, through the
same detector and the same false-alarm rate, so no property of the front end can
account for a difference between them: a clutter or threshold deficiency is common
to all three and would move them together. What differs is where each target
\label{sec:assoc:kpi}
flies, and the two that degrade are the two that fail the screens above, each on a
different one. The well-sited target meets the
\SIrange{\MTkpiLo}{\MTkpiHi}{\metre} horizontal position accuracy identified for
the UAV use case in the 3GPP feasibility study~\cite{3gpp_tr22837} --- with the
caveat of Section~\ref{sec:intro} that this is a study report, and that the range
is a use-case requirement for a sensing service rather than a conformance limit
for a sensor, so meeting it on one target is evidence about siting and not a
certification. Table~\ref{tab:mt:kpi} states the comparison in full, including the
two use-case KPIs this evaluation does not measure at all.

\paragraph{Scoring on nearest-consistent pairings}
\label{sec:assoc:attribution}
Attributing tracks to targets is itself a per-CPI one-to-one assignment, made at
a \SI{100}{\metre} radius, and it has a failure mode that falls entirely on one
target here. When two tracks chase the same UAV only one of them can be assigned
to it, and the other goes to whichever \emph{other} UAV is still inside the
radius. On a scene whose targets are \SIrange{33}{100}{\metre} apart in the
ground plane that radius
is wider than the scene, so the displaced track lands on a real UAV rather than
being discarded, and its distance to \emph{that} UAV enters that UAV's row.
\MTassocForced\% of per-CPI assignments across the baseline campaign are of this
kind. We therefore score each target only over pairings in which the assigned UAV
is also the one the track is nearest to; on the pinned seed the two tracks
populating UAV-B's row under the permissive rule sit \SI{18}{\metre} and
\SI{76}{\metre} from \emph{UAV-A} against \SI{52}{\metre} and \SI{83}{\metre}
from UAV-B, and the second is nearest to nothing at all.

Scored at the assignment radius the same three targets read
\SI{\MTaPosRmseAR}{\metre}, \SI{\MTbPosRmseAR}{\metre} and
\SI{\MTcPosRmseAR}{\metre}. The difference is confined to UAV-B: UAV-A and UAV-C
move by \SI{0.1}{\metre} and \SI{0.2}{\metre}, and the \MTposSpread$\times$
spread that carries the argument above is materially unchanged, so nothing in
this section rests on the choice. UAV-B loses a third of its figure, along with
\MTbContAR\% $\rightarrow$ \MTbCont\% of its continuity and one of the
\MTcampN{} runs these rows are pooled over, in which no track is nearest to
UAV-B at all. What the correction buys is a coherent diagnosis rather than a
better number, seen in the range/cross-range split of
Table~\ref{tab:mt:tracking}.

\paragraph{Scope of the cross-range attribution}
At $N=\MTcampBatch$ UAV-B's cross-range error was nearly twice its range error
and we read that as an azimuth failure. At $N=\MTcampN$ the two components are
equal (\MTbCrossRmse~m against \MTbRangeRmse~m) and the reading does not survive:
per seed, UAV-B's cross-range error is below \SI{10}{\metre} on six of
\MTcampNb{} and above \SI{40}{\metre} on \MTbCrossOutliers, and at the smaller
sample those \MTbCrossOutliers{} were half of it. We therefore withdraw the
azimuth attribution \emph{for UAV-B} and report its error as dominated by
neither component.
The attribution belongs to UAV-C, where it is not sample-dependent: cross-range
exceeds range on every one of its \MTcampNc{} scoreable seeds
(\MTcCrossRmse~m against \MTcRangeRmse~m, medians
\MTcCrossRmseMed{} and \MTcRangeRmseMed), and doubling the sample strengthened it.
UAV-C is also the target whose grazing geometry the azimuth screen of
Section~\ref{sec:assoc:birth} predicts should fail, so the surviving case is the
one the mechanism predicted.
Both readings leave UAV-B outside the study range, so no verdict in
Table~\ref{tab:mt:kpi} turns on it either.

Relaxing the one-to-one constraint instead is not the alternative it appears to
be: it would let one estimate be credited to two targets at once, which is a
worse error in the opposite direction and, on the diagonal geometry of
Section~\ref{sec:assoc:heading}, would report two tracked UAVs where the detector
delivered one. The fragmentation that creates the displaced track in the first
place is a tracker property, not a scoring one --- neither rule addresses it, and
it is what Section~\ref{sec:assoc:birth} attempts to control.

\begin{table}[!tbp]
  \centering
  \caption{Achieved performance against the UAV use-case KPIs of the 3GPP feasibility study~\cite{3gpp_tr22837}. Per-target rows are unbounded and scored on nearest-consistent attributions (Section~\ref{sec:assoc:birth}); the pooled row is the set-level matcher's own \SI{20}{\metre}-gated figure, not the sum of the three. The study is a \emph{study report} --- use-case requirements, not conformance limits (Section~\ref{sec:intro}). $^\dagger$UAV-B is the one row the attribution rule moves materially (Section~\ref{sec:assoc:birth}); UAV-A and UAV-C move by less than \SI{0.25}{\metre}. All three miss the range. The parenthesised labels name the screen each target fails; UAV-B's read ``fails azimuth'' until $N=\MTcampN$, at which point its two error components became equal and that attribution was withdrawn (Section~\ref{sec:assoc:birth}) --- the label follows the text rather than the other way round.}
  \label{tab:mt:kpi}
  \footnotesize
  \setlength{\tabcolsep}{4pt}
  \begin{tabular}{@{}llc@{}}
    \toprule
    Key performance indicator (study range) & achieved & meets \\
    \midrule
    \multicolumn{3}{@{}l}{\emph{Measured: horizontal position, \SIrange{\MTkpiLo}{\MTkpiHi}{\metre}}} \\
    \quad UAV-A (near broadside)  & \SI{\MTaPosRmse}{\metre}  & yes \\
    \quad UAV-B (neither dominates) & \SI{\MTbPosRmse}{\metre}\rlap{$^\dagger$}  & no  \\
    \quad UAV-C (fails elevation) & \SI{\MTcPosRmse}{\metre}  & no  \\
    \quad pooled over three       & \SI{\MTrmseUngated}{\metre} & no \\
    \midrule
    \quad Altitude (same range) & \MTaltRmse~m & at edge \\
    \midrule
    \multicolumn{3}{@{}l}{\emph{Study-report requirements this evaluation does not measure}}\\
    \quad Velocity, \SIrange{1}{10}{\metre\per\second} & not evaluated & --- \\
    \quad Latency, $<\SI{20}{\milli\second}$           & not evaluated & --- \\
    \bottomrule
  \end{tabular}
\end{table}

The practical reading is that this sensor's accuracy is governed by geometry
between the node and the target rather than by detector performance, and that a
pooled figure over targets at unequal standoff describes the siting of the
scenario more than the capability of the array. We therefore report per-target
accuracy alongside the pooled value throughout, and treat the analytic screens as
a deployment tool rather than only an analysis one.

An explicit escape---allowing a measured elevation disagreement of more than
\MTbirthElSep$^\circ$ to disqualify an incumbent from vetoing a birth, that
incumbent only---was designed for the two-dimensional regime, where it was the
only way to distinguish stacked targets. Retained on top of the
three-dimensional filter it is no longer a net gain. It buys coverage, raising it
from \MTcovNoElSep\% to \MTcovElSep\%, and it pays for that coverage in identity:
IDF1 falls from \MTidfNoElSep{} to \MTidfElSep, identifiers per target rise from
\MTidsNoElSep{} to \MTidsElSep, and identifier switches rise from
\MTidswNoElSep{} to \MTidswElSep. All four differences separate at $N=\MTcampN$
($p \le 0.004$ on paired seeds, exact sign-flip).

Whether it helps the targets it was meant to protect is resolvable
 over the $N=\MTcampN$ runs, on both weak targets and in the same direction. The served
fraction rises from \MTescBservedOff\% to \MTescBservedOn\% for UAV-B
($p=\MTescBservedP$) and from \MTescCservedOff\% to \MTescCservedOn\% for UAV-C
($p=\MTescCservedP$), with scene coverage rising \MTescCovOff\% to
\MTescCovOn\% ($p=\MTescCovP$). The escape does what it was designed to do: the
targets birth inhibition was suppressing are the targets it recovers.

\paragraph{Threshold dependence of the coverage--identity trade}
A mechanism characterized at one threshold invites the question of whether its
behavior is discontinuous near that value. Sweeping it from \ang{\MTescSweepLo} to
\ang{\MTescSweepHi} over the same runs and the same detections, every quantity
moves \emph{monotonically} and nothing discontinuous happens anywhere near the
shipped value: lowering the threshold admits more escapes and raises coverage as
far as \MTescSweepCovLo\%, at \MTescSweepIdswLo{} identity switches per transit,
while raising it converges smoothly back onto the disabled arm.

The informative part is not the range but its uniformity. Coverage bought per
additional identity switch is \MTescRateMin{} to \MTescRateMax~pp across that
whole span --- mean \MTescRateMean~pp, and no trend with the threshold. The
threshold therefore chooses \emph{how much} of the trade to take, not how
favorable it is: there is no setting at which coverage becomes cheap, and none
at which it becomes disproportionately expensive.

We therefore disable it in the configuration all other results use, and state the
choice as a choice: this work prioritizes identity, because a downstream C2
consumer is handed identifiers and cannot use a track that changes identity eleven
times per transit. The sweep is what makes that defensible rather than tuned ---
the escape is not disabled because \ang{16} happened to be a poor operating point,
but because \emph{every} operating point charges the same rate. A deployment that
valued coverage over identity would reasonably set it the other way, and the
figures above are what it would gain and lose.

One way to keep the coverage without the identity cost is to demand
persistence: require an incumbent to be contradicted on $M$ \emph{consecutive}
intervals before it loses its veto, since the escape threshold sits at
\MTescSigmaK$\sigma$ of a noisy observable and fires at the tail rate on a single
look, whereas a genuinely distinct target should contradict every interval. We
implemented it, and it does not work. At $M=2$ the escape retains
\MTpersistCovKept\% of its coverage gain but \MTpersistIdswKept\% of its
identifier-switch cost, so the exchange rate moves against it; at $M=3$ every
scene-level metric is identical to disabling the escape outright, because no track
in this scene is contradicted on three consecutive intervals. The premise fails: a
distinct target contradicts on every interval only if it is \emph{detected} on
every interval, and the weak targets are held on
\SIrange{\MTpersistHoldLo}{\MTpersistHoldHi}{\percent} of them. The requirement
conditions on detection continuity rather than veracity, and detection continuity
is poorest for exactly the targets the escape recovers---so it removes the true
escapes before the false ones.

This does not resolve identity. Section~\ref{sec:cadence:residual} reports the
residue---duplicate identifiers held concurrently on one target once the cadence
artifact is removed---as an open problem in birth admission.

\subsection{Association Algorithm and Birth Geometry}
\label{sec:assoc:ablation}

The binding limit here is identity, not
detection: targets are detected far more often than they are tracked, and the
surplus is concurrent rather than sequential --- two identifiers held on one
target at the same time (Section~\ref{sec:cadence:residual}). Closing it needs
track management that absorbs a multi-detection interval without re-spawning, and
association that scores a detection against the whole hypothesis set rather than
one track at a time. We have measured the second of those two: joint
probabilistic association (JPDA), which weights a contested detection across all
tracks that gate it, re-scored offline on the same recorded detections with the
association rule as the only difference, paired within each run, $N=\MTjpdaStraightN$
per geometry. \emph{It does not close the identity gap on the reference
geometry.} Identifiers per target move \MTjpdaStraightIdsPda{} to
\MTjpdaStraightIdsJpda{} ($p=\MTjpdaStraightIdsP$) and identity switches
\MTjpdaStraightIdswPda{} to \MTjpdaStraightIdswJpda{}
($p=\MTjpdaStraightIdswP$); what moves instead is scene coverage,
\MTjpdaStraightCovPda\% to \MTjpdaStraightCovJpda\% in
\MTjpdaStraightCovWins{} of \MTjpdaStraightN{} runs ($p=\MTjpdaStraightCovP$),
paid for in localization --- the GOSPA localization term rises
\MTjpdaStraightGospaLocPda{} to \MTjpdaStraightGospaLocJpda{} in
\MTjpdaStraightGospaLocLosses{} of \MTjpdaStraightN{} runs
($p=\MTjpdaStraightGospaLocP$), leaving total GOSPA unchanged
($p=\MTjpdaStraightGospaP$). Most of that rise is cost \emph{re-accounted}
rather than accuracy lost: GOSPA charges an unmatched truth to the missed term
and a matched-but-distant one to the localization term, so intervals where the
two configurations report a different number of estimates carry
$\MTjpdaStraightLocDiffCard$ of localization against
$\MTjpdaStraightMissDiffCard$ of missed, while on the
\MTjpdaStraightSameCardPct\% of intervals with equal cardinality the
localization penalty is only $\MTjpdaStraightLocSameCard$. On the alternative heading, where two targets
contend for one detection far more often, the same change acts on the terms it
is supposed to: total GOSPA \MTjpdaDiagGospaPda{} to \MTjpdaDiagGospaJpda{} and
horizontal RMSE \MTjpdaDiagRmsePda{} to \MTjpdaDiagRmseJpda{}~m in
\MTjpdaDiagGospaWins{} of \MTjpdaDiagN{} runs ($p=\MTjpdaDiagGospaP$), the false
term \MTjpdaDiagGospaFalsePda{} to \MTjpdaDiagGospaFalseJpda{}
($p=\MTjpdaDiagGospaFalseP$) and identifiers per target \MTjpdaDiagIdsPda{} to
\MTjpdaDiagIdsJpda{} ($p=\MTjpdaDiagIdsP$), at coverage that does not separate
($p=\MTjpdaDiagCovP$). Joint association is therefore a
\emph{geometry-dependent trade rather than a fix}, and the identity surplus this
paper reports is not the term it acts on.

The track management above \emph{is} that term, and re-scoring the same
detections isolates which part of it. A replay ledger over the two campaigns
records no confirmed track ever being dropped ---
\MTbgStraightDropsConf{} and \MTbgDiagDropsConf{} per run on the two headings ---
so the surplus cannot be death-and-rebirth, and must be births that survive to
confirmation while an incumbent still holds the target. That makes the birth
\emph{geometry} the lever rather than the association rule. Widening the region
in which an existing track inhibits a new one from twice the association gate to
four times, again with everything else fixed and paired within each run
($N=\MTbgStraightN$), moves identifiers per target
\MTbgStraightIdsKtwo{} to \MTbgStraightIdsKfour{} on the reference heading
($p=\MTbgStraightIdsP$) and \MTbgDiagIdsKtwo{} to \MTbgDiagIdsKfour{} on the
alternative ($p=\MTbgDiagIdsP$), with the GOSPA false term falling
\MTbgStraightGospaFalseKtwo{} to \MTbgStraightGospaFalseKfour{}
($p=\MTbgStraightGospaFalseP$) and \MTbgDiagGospaFalseKtwo{} to
\MTbgDiagGospaFalseKfour{} ($p=\MTbgDiagGospaFalseP$). It is a trade rather than
a free fix, and in the opposite direction to JPDA's: scene coverage falls
\MTbgStraightCovKtwo\% to \MTbgStraightCovKfour\%
($p=\MTbgStraightCovP$) and \MTbgDiagCovKtwo\% to \MTbgDiagCovKfour\%
($p=\MTbgDiagCovP$), and the loss is concentrated on the two targets whose
returns are weakest --- a wider inhibition region withholds an identifier
exactly where evidence for a second one is thinnest. Which point to operate at is
therefore a property of the consumer, not of the tracker: a fusion node that must
hold a stable identity across a transit values the first column, and a coverage
figure values the second. Every other number in this paper is scored at
the shipped setting, twice the gate; the comparison here is the only place the
wider one appears.

\begin{table}[t]
\caption{Per-target tracking accuracy over $N=\MTcampN$ seeded transits ($N=\MTcampNb$ for UAV-B and $N=\MTcampNc$ for UAV-C, which lack a scoreable attribution in some runs). Rows are scored on nearest-consistent attributions (Section~\ref{sec:assoc:birth}) and are unbounded (not at the \SI{20}{\metre} gate), so they exceed the pooled RMSE of Table~\ref{tab:mt:basecmp} rather than contradicting it. UAV-B's cross-range mean is not representative of its distribution: \SIrange{\MTbCrossRmseLo}{9.2}{\metre} on six of \MTcampNb{} seeds and \SI{\MTbCrossRmseHi}{\metre} on the other two (median \SI{\MTbCrossRmseMed}{\metre}). Primary-identity RMSE is listed for comparison (Section~\ref{sec:assoc:metrics}).}
\label{tab:mt:tracking}
\centering
\begin{tabular}{lrrr}
\toprule
Target & RCS (dBsm) & $\Delta R$ RMSE (m) & Cross-range RMSE (m) \\
\midrule
UAV-A & \MTaRcs & \MTaRangeRmse & \MTaCrossRmse \\
UAV-B & \MTbRcs & \MTbRangeRmse & \MTbCrossRmse \\
UAV-C & \MTcRcs & \MTcRangeRmse & \MTcCrossRmse \\
\midrule
\multicolumn{2}{l}{Primary-identity RMSE} & \multicolumn{2}{r}{\MTprimaryRmse} \\
\bottomrule
\end{tabular}
\end{table}

\subsection{Sensitivity to Target Heading: a Diagonal Transit}
\label{sec:assoc:heading}

All results above are obtained on one set of headings, in which the three UAVs fly
parallel south--north transits. That arrangement is favorable to association:
each target sweeps the array on its own bearing schedule, so the three are
angularly well separated for most of the transit. To test whether the conclusions
survive a geometry that removes that separation, we repeat the campaign with UAV-B
rotated to a \ang{45} south-to-north-east diagonal at the same speed, altitude and
cross section, with its excursion clamped symmetrically about the gNB boresight.
UAV-A and UAV-C are configured identically in the two scenes, so the manipulation
is one \emph{configured} variable. It does not follow that the two rendered
scenes differ in one \emph{measured} variable, and the result below is where that
distinction is paid for.

The geometric consequence is measured from the ground truth of the two rendered
traces before any run, and it is not the expected one. \emph{Sensitivity} barely
moves: UAV-B's zero-Doppler notch occupancy rises only from \MThdgBaseNotch\% to
\MThdgDiagNotch\% --- both well short of the \SI{25}{\percent} occupancy at which
a target becomes unusable --- and its reportable fraction falls only from
\MThdgBaseReport\% to \MThdgDiagReport\%. Both are sensitivity-side quantities---whether a lone
target would be reported at all---and it matters for what follows that the
prediction was formed from them. What the geometry changes is angular
\emph{separability}, the input to contention. On the
diagonal, UAV-B co-moves with UAV-A in bearing rather
than crossing it: the pair's median azimuth separation collapses from
\ang{\MThdgBaseDaz} to \ang{\MThdgDiagDaz} and its median elevation separation
from \ang{\MThdgBaseDel} to \ang{\MThdgDiagDel}. Measured against the angle
uncertainties the tracker assumes, the A--B pair is separated by neither azimuth
nor elevation in \MThdgDiagAmb\% of intervals on the diagonal, against
\MThdgBaseAmb\% on the baseline; range separation still resolves
\MThdgDiagAmbRange\ percentage points of that, and the remainder is left to the
association gate.

Range separation resolves less of it than the waveform allows, and the reason is
the detector rather than the bandwidth. The pair is separated by more than the
\SI{\MTrangeRes}{\metre} bistatic range cell in every interval of both scenes, so
nothing here is at the resolution limit; but the deployed non-maximum suppression
discards the weaker of two peaks within \SI{\MTnmsRadius}{\metre} of each other
in range, which is \num{2.1} times that cell. Scored on the radius the detector
actually uses, the diagonal merges the A--B pair into a single peak in
\MThdgDiagNmsMerged\% of intervals against \MThdgBaseNmsMerged\% on the baseline.
That suppression radius is a tuning choice inherited from the single-target
configuration, where its only job was to stop one target producing several peaks
(Table~\ref{tab:params}); on this geometry it is the mechanism by which two
targets become one detection, and it bounds what any downstream association can
recover.

The diagonal therefore also removes the elevation escape that
Section~\ref{sec:upa} relies on: the fraction of intervals in which elevation is
the deciding discriminant for this pair falls from \MThdgBaseElDec\% to
\MThdgDiagElDec\%, because UAV-B descends into UAV-A's elevation band instead of
sitting above it. This is the adverse case for the planar array's contribution,
and it is the reason to run it.

The comparison was registered in these terms before the campaign was run, with the
outcome that would contradict it named in advance: if \emph{association quality}
is the binding constraint, then a geometry that leaves sensitivity almost
unchanged while destroying angular separability should degrade the identity
metrics substantially and per-target availability hardly at all; the opposite
pattern would indicate the attribution is wrong.

\emph{Result.} Over $N=\MThdgDiagCampN$ seeded transits of the diagonal scene,
scored on the sensor timeline with the identical configuration and aggregation as
the baseline, \emph{the opposite pattern is what we observe}
(Table~\ref{tab:mt:heading}). The contested-cell fraction rises from
\MTcontested\% to \MThdgDiagContested\%, so nearly every interval now has two
targets competing for one detection, and per-target availability falls.

The drop is not confined to the target that was moved, which bounds what this experiment attributes. UAV-B, the only target
whose trajectory changed, drops \MThcPdBDrop~percentage points
($|t| = \MThcPdBT$) --- but so do both targets we did not touch, UAV-A by
\MThcPdADrop~points ($|t| = \MThcPdAT$) and UAV-C by \MThcPdCDrop{}
($|t| = \MThcPdCT$). The manipulation therefore acted on the scene rather than on
one UAV: raising contention costs every target, by a similar amount. That is
consistent with contention being the binding mechanism, and it is stronger
evidence for it than a UAV-B-only drop would have been. It also means the
control we had designed---unchanged targets hold still---did \emph{not} hold, so
this comparison cannot separate the heading change from any other difference
between two separately rendered scenes. We report it as such.

The identity metrics split. IDF1 does not move at all---\MThdgDiagIdfOne{}
against \MTidfOne{} ($|t| = \MThcIdfT$)---and neither does the set-level distance
(GOSPA \MThdgDiagGospa{} against \MTgospa, $|t| = \MThcGospaT$). But the two
fragmentation-sensitive measures degrade clearly: identifiers per target rise from
\MTidentifiers{} to \MThdgDiagIdentifiers{} ($|t| = \MThcIdsT$) and switches per
transit from \MTidsw{} to \MThdgDiagIdsw{} ($|t| = \MThcIdswT$), both unpaired
across the two scenes at $N=\MTcampN$. The diagonal geometry therefore costs
\emph{track continuity}
without costing \emph{assignment correctness}: the tracker keeps putting the right
detection on a target, and keeps losing and re-minting the identifier it does so
under. IDF1 and GOSPA are comparatively insensitive to that, which is why they are
quoted alongside the identifier counts rather than instead of them.

\begin{table}[!tbp]
  \centering
  \caption{The heading experiment, registered prediction against outcome. Both
  arms $N=\MThdgDiagCampN$, identical configuration and aggregation, scored on the
  sensor timeline; $|t|$ is Welch's, unpaired, because the two arms necessarily
  fly separately rendered scenes. The prediction was fixed before the diagonal
  campaign ran (Section~\ref{sec:assoc:heading}).}
  \label{tab:mt:heading}
  \footnotesize
  \setlength{\tabcolsep}{3.5pt}
  \begin{tabular}{@{}lrrrl@{}}
    \toprule
    & Baseline & Diagonal & $|t|$ & Predicted \\
    \midrule
    \multicolumn{5}{@{}l}{\emph{Manipulation} (ground truth, pre-run)}\\
    A--B azimuth sep.\ (\si{\degree}) & \MThdgBaseDaz & \MThdgDiagDaz & --- & collapse \\
    A--B elevation sep.\ (\si{\degree}) & \MThdgBaseDel & \MThdgDiagDel & --- & collapse \\
    Neither-separated (\%) & \MThdgBaseAmb & \MThdgDiagAmb & --- & rise \\
    UAV-B notch occ.\ (\%) & \MThdgBaseNotch & \MThdgDiagNotch & --- & $\approx$ \\
    UAV-B reportable (\%) & \MThdgBaseReport & \MThdgDiagReport & --- & $\approx$ \\
    \midrule
    \multicolumn{5}{@{}l}{\emph{Availability}} \\
    Contested cells (\%) & \MThcContestedBase & \MThcContestedDiag & \MThcContestedT & --- \\
    UAV-A avail.\ (\%, control) & \MThcPdABase & \MThcPdADiag & \MThcPdAT & $\approx$ \\
    UAV-B avail.\ (\%, \emph{moved}) & \MThcPdBBase & \MThcPdBDiag & \MThcPdBT & $\approx$ \\
    UAV-C avail.\ (\%, control) & \MThcPdCBase & \MThcPdCDiag & \MThcPdCT & $\approx$ \\
    \midrule
    \multicolumn{5}{@{}l}{\emph{Identity}} \\
    IDF1 & \MThcIdfBase & \MThcIdfDiag & \MThcIdfT & degrade \\
    GOSPA (\si{\metre}) & \MThcGospaBase & \MThcGospaDiag & \MThcGospaT & degrade \\
    Identifiers / target & \MThcIdsBase & \MThcIdsDiag & \MThcIdsT & degrade \\
    Identity switches & \MThcIdswBase & \MThcIdswDiag & \MThcIdswT & degrade \\
    \bottomrule
  \end{tabular}

  \vspace{2pt}
  \parbox{\columnwidth}{\scriptsize The prediction fails in both blocks and in
  opposite directions. The two metrics it named---IDF1 and GOSPA---do not move,
  while availability, which it expected to hold, falls for \emph{all three}
  targets: furthest for UAV-C, which was not moved. What degrades instead is
  identifier maintenance, which the prediction did not mention.}
\end{table}

\subsection{Validation Controls}
\label{sec:assoc:guards}

Multi-target results of this kind admit two failure modes that a headline number
will not expose, so both are run as explicit controls.

\paragraph{Over-merge}
Birth inhibition and probabilistic association both act to suppress surplus
identifiers. Taken too far, they achieve a perfect identifier count by merging
genuinely distinct targets into one track---an outcome that improves every
identity metric while destroying the capability being claimed. The control scores
identifier counts on the CPIs where the three targets are known to be resolvable
in range or Doppler: collapsing toward a single track there is a failure, not a
pass. The measured value is \MTovermergeIds.

\paragraph{Target-absent false tracks}
Conversely, birth inhibition could suppress spurious tracks simply by suppressing
births in general, which would also suppress legitimate ones. The control replays
a target-absent channel through the identical configuration; the false-track term
must remain near zero while the true-track term on the populated scenario is
undegraded. This control is load-bearing here rather than routine, because
raising the detector's capacity (Section~\ref{sec:mt:rawcap}) admitted a
\MTfalseFrac\% false-detection load that the previous saturated buffer had been
discarding silently. The measured false-track term is \MTabsentFalse.

\section{Measurement-Model Corrections}
\label{sec:measmodel}

A single-target evaluation exercises one point in the observation space. Placing
three targets at differing altitudes and bearings exercised several, and in doing
so exposed three defects --- two in the measurement model, one in its noise
assumptions --- that a single well-placed target cannot reveal. All three are
reported here because all three were silent: none raised an error, and each
produced a plausible number.

\paragraph{Corrections active in the reported results}
Table~\ref{tab:measmodel:ablation} gives each correction, the before/after
measurement that establishes it, and---the column that matters for reading
Tables~\ref{tab:mt:basecmp}--\ref{tab:mt:heading}---whether it was active in the
campaign those tables are scored from.

Two of the three are active; the bistatic range-bias constant is not, and we correct an earlier statement here: it is listed in
Table~\ref{tab:params} as a configuration parameter, which overstates it. The
constant is set by the campaign's environment file but is not forwarded by the
container orchestration to the process that reads it, so it never took effect in
any run reported in this paper. The consequence is:
\emph{every bistatic range figure in this paper carries the uncorrected
\SI{\MTrangeBias}{\metre} under-read}. Measured on a single campaign run, the
median residual between the reported range and the true total bistatic path is
\SI{-7.3}{\metre}, which is the uncorrected value and not the corrected one.

\begin{table}[t]
  \caption{The three measurement-model corrections, and whether each was active
  in the campaign the results are scored from. The range-bias constant was
  configured but never reached the estimator; it is characterized here, not
  applied.}
  \label{tab:measmodel:ablation}
  \footnotesize
  \setlength{\tabcolsep}{3pt}
  \begin{tabular}{@{}lccc@{}}
    \toprule
    Correction & Before & After & Live? \\
    \midrule
    Cone angle vs.\ ground bearing            &                    &                  &     \\
    \quad UAV-B azimuth residual (\si{\degree})   & \MTbAzResidGround  & \MTbAzResidCone  & yes \\
    \quad residual SD (\si{\degree})              & \MTbAzSdGround     & \MTbAzSdCone     & yes \\
    \addlinespace
    Elevation variance floor                  &                    &                  &     \\
    \quad UAV-A altitude SD (m)               & \MTaZsdBefore      & \MTaZsdAfter     & yes \\
    \addlinespace
    Bistatic range bias (\SI{\MTrangeBias}{\metre}) &              &                  &     \\
    \quad median range residual (m)           & \MTrangeResidBefore & \MTrangeResidAfter & \textbf{no} \\
    \bottomrule
  \end{tabular}
\end{table}

\subsection{Angle of Arrival as a Direction Cosine}
\label{sec:measmodel:cone}

The receive array is a uniform planar array whose horizontal axis measures the
direction cosine of the arrival along that axis---the cone angle
$\arcsin(\Delta y / R_{\mathrm{slant}})$, where $R_{\mathrm{slant}}$ is the
three-dimensional range from array to target. The filter's observation function
predicted instead the ground-plane bearing
$\operatorname{atan2}(\Delta y, \Delta x) = \arcsin(\Delta y / R_{\mathrm{ground}})$.
The two differ by a factor of $\cos(\text{elevation})$ and coincide exactly when
the target is at the array's own height or at broadside---which is why two
single-target studies did not encounter it: their canonical target is near
broadside.

The discriminating measurement is the residual at ground truth, where the only
remaining error is the model's own. Of the three targets, only UAV-B is
simultaneously elevated and off-broadside, and only UAV-B shows the effect: its
median azimuth residual is \MTbAzResidGround$^\circ$ under the ground-bearing
model and \MTbAzResidCone$^\circ$ under the cone model, with the residual standard
deviation falling from \MTbAzSdGround$^\circ$ to \MTbAzSdCone$^\circ$. UAV-A and
UAV-C are unchanged to within a fraction of a degree under either model, exactly
as the $\cos(\text{elevation})$ dependence predicts. This is a correctness fix: the legacy expression is wrong for any target that has both
height and an off-broadside bearing, which in deployment is most of them.

Correcting the model allows the angular measurement covariance to be set from the
data rather than inflated to absorb a systematic error. Re-estimated from the
truth-referenced residual spread, the angular and
range standard deviations fall to \MTsigmaAzMeas$^\circ$ and \MTsigmaRMeas~m from
the inherited single-target values.

\subsection{Correlation of Elevation Errors Across CPIs}
\label{sec:measmodel:elcorr}

The vertical channel deserves separate treatment because its error structure
violates an assumption the filter makes implicitly. A Kalman update shrinks the
altitude variance in proportion to the number of independent measurements
absorbed. Measured at ground truth, the elevation residual is not independent
across CPIs: it carries a per-target static offset and, on two of the three
targets, a lag-one autocorrelation of approximately \MTelRho{} that does not decay
with lag---the signature of a random constant per track rather than of white
noise. A constant does not average away, so the altitude error never falls below
its single-look value while the reported variance continues to shrink.

The arithmetic closes. The single-look altitude standard deviation implied by the
measured elevation error and the observation geometry is \MTzSingleLook~m across
the three targets, against a reported \MTzReported~m; the resulting ratios match
the observed over-confidence. Constraining the altitude variance to the value one
elevation measurement supports removes the discrepancy, and---because a larger
variance keeps the elevation gain from collapsing onto an early estimate---also
improves accuracy, reducing UAV-A's altitude error standard deviation from
\MTaZsdBefore~m to \MTaZsdAfter~m.

\section{Separating Sensor Limits from Evaluation Limits}
\label{sec:cadence}

A digital twin that cannot run its scenario in real time does not merely take
longer to produce a result: it can change the result. This section isolates that
effect, which on this testbed dominates every continuity and identity figure
reported for multiple targets. The distinction it draws---between what the
\emph{sensor} could not do and what the \emph{workstation} could not do---is the
difference between a negative result and a positive one.

\subsection{Emulator Throughput Below Real Time}
\label{sec:cadence:dilation}

The gNB assembles one sensing CPI from a fixed number of SRS bursts, which fixes
the CPI period at \MTcpiPeriod~s of scene time regardless of how fast the host
executes. On the multi-target configuration the channel emulator convolves
\MTtapsPerUav{} taps across \MTnRxPorts{} receive ports per batch, against a
\MTrtBudget~$\mu$s real-time budget; the measured cost is \MTbatchUs~$\mu$s, a
factor of \MToverBudget{} over. The host therefore delivers slots more slowly than
the scene evolves, and consecutive CPIs arrive \MTwallGap~s apart on the wall
clock instead of \MTcpiPeriod~s.

That this is dilation rather than under-sampling is directly verifiable from the
ground truth. Between consecutive detection-bearing CPIs the three targets advance
\MTaStep, \MTbStep{} and \MTcStep~m respectively---in each case exactly the
distance the target covers in \MTcpiPeriod~s at its own speed. The sensor sampled
the scene at its design rate; only the wall clock was stretched, by
\MTdilation$\times$.

Stated as rates, so that neither can be quoted for the other: the sensor's
\emph{design} cadence is $1/\MTcpiPeriod~\mathrm{s} = \SI{\MTrateDesign}{\hertz}$,
and the \emph{emulator throughput} actually achieved is
$1/\MTwallGap~\mathrm{s} = \SI{\MTrateWall}{\hertz}$. Every cadence figure in this
paper is one of those two and is labeled as such. \emph{Neither has been measured on radios}: the design rate follows from the SRS schedule and is not a
demonstrated capability, and the throughput rate is a property of this
workstation. Any comparison against an operational refresh-rate requirement must
be made against a hardware measurement that this paper does not contain.

\subsection{Two Wall-Clock-Dependent Tracker Constants}
\label{sec:cadence:constants}

Most of the processing chain is indifferent to this, because it is driven by
samples rather than by time. Two quantities are not, and both sit in the
evaluation path rather than the sensing path.

The first is track publication lifetime. A published track is a key with a
\MTttl~s expiry, refreshed whenever a detection updates the track. Against a
\MTcpiPeriod~s CPI period that bridges the interval between detections many times
over; against a \MTwallGap~s dilated gap it expires long before the next
detection. The second is the coverage denominator itself. Coverage is scored at
the recorder's polling instants, so a tracker that is correct at every single
detection still scores only $\min(\text{gap},\text{TTL})/\text{gap}$ for a
publication time-to-live (TTL): on these
runs a \emph{ceiling} of \MTceilingWall\%, not 100\%.

The consequence for the standard metric set invalidates a figure that is
otherwise reported without qualification. A target is conventionally ``mostly
tracked'' if it is tracked for more than 80\% of its life.
At a \MTceilingWall\% ceiling that is unreachable by construction, so a
mostly-tracked count of zero carries no information about the tracker at all. Any
such count must be reported against its ceiling or not at all.

\subsection{Replay on the Sensor's Timeline}
\label{sec:cadence:simtime}

Because the detections and the ground truth are both recorded, the evaluation can
be repeated on the timeline the sensor actually sampled: the same detections, in
the same order, re-spaced at the \MTcpiPeriod~s CPI period, driving the same
tracker with no parameter changed. Nothing about the sensing is altered---no
detection is added, removed or moved relative to its CPI---and the tracker's own
constants are left exactly as deployed. Only the wall-clock dilation is removed.

Table~\ref{tab:cadence} gives the comparison over $N=\MTcampN$ seeded transits.

Re-timing the evaluation does not change the basis of comparison, because \emph{the detection stream is independent of the wall clock by construction}. The channel is a deterministic ray-traced replay, so the sequence
of echoes is fixed before the run starts; the sensing chain completes in
\MTpipelineMs~ms per CPI on average and \MTpipelineMax~ms at worst, inside the
\MTcpiWindowMs~ms window, so no CPI was dropped for want of host time. Re-timing cannot add a detection, remove
one, or move one relative to its CPI; it changes exactly two tracker constants
that were denominated in the wrong units.

\paragraph{Interpretation of the sensor-timeline figures}
They are an \emph{artifact-removal analysis}: the same detections scored with the
two mis-denominated constants corrected, which isolates the tracker's behavior
from the emulator's execution rate. They are \emph{not} a bound on what a
real-time deployment would achieve. A replay is
deterministic and unloaded by construction. A deployment adds at least: RF
front-end conversion and its impairments, scheduler contention when the sensing
stage shares a host with a live RAN, queueing and retransmission on a real E2
link, CPIs dropped under transient load, and interference that no ray-traced
replay contains. Any of these can move a result in either direction --- dropped
CPIs lengthen detection gaps, while a real receiver's noise realizations differ
from the injected ones --- so the correct statement is that the replay removes a
known evaluation artifact, not that it brackets deployment performance.

What \emph{is} measured is the part of the budget this testbed executes. Over
\MTlatRecords{} published track records of the live E2 path, the PHY sensing
stage completes in \MTpipelineMs~ms per CPI on average (\MTpipelineMax~ms at
worst), gNB$\rightarrow$RIC$\rightarrow$xApp transport costs
\SI{\MTlatEtwoMed}{\milli\second} at the median (\SI{\MTlatEtwoPnf}{\milli\second} at
the 95th percentile), and the filter update itself
\SI{\MTlatEkfMed}{\milli\second} (\SI{\MTlatEkfMax}{\milli\second} at worst) ---
about \SI{\MTlatTotalMed}{\milli\second} end to end inside a
\MTcpiWindowMs~ms CPI window, dominated by the sensing stage rather than by
transport or filtering. Those three components are wall-clock measurements from
the live run, not replay values; the replay reports zero transport latency by
construction, which is why the budget is quoted from the live path. Not in this
budget, and not measured anywhere here: RF conversion, host contention from a
co-resident RAN, and anything on the C2 side of the adapter.

\begin{table}[t]
\caption{The same detections and estimator scored on the dilated wall clock and on the sensor's own timeline; only poll spacing differs. Mostly / partly / lost are counts over the \MTmotaTargetRuns{} target-runs ($N=\MTcampN$ seeds $\times$ \MTnTargets{} targets).}
\label{tab:cadence}
\centering
\begin{tabular}{lrr}
\toprule
 & Wall clock & Sensor timeline \\
\midrule
Coverage ceiling            & \MTceilingWall\%   & \MTceilingSim\% \\
Coverage                    & \MTcovWall\%       & \MTcovSim\% \\
Fragmentations per transit  & \MTfmWall          & \MTfmSim \\
IDF1                        & \MTidfWall         & \MTidfSim \\
Primary-identity RMSE (m)   & \MTprimaryWall     & \MTprimarySim \\
Horizontal RMSE (m)         & \MTrmseWall        & \MTrmseSim \\
Identifiers per target      & \MTidsPerTgtWall   & \MTidsPerTgtSim \\
Mostly tracked / partly / lost & \MTmotaWall     & \MTmotaSim \\
\bottomrule
\end{tabular}
\end{table}

\paragraph{Both columns are produced by replay}
The wall-clock column is not the tracker output the live run recorded. It is the
same replay as the sensor column with the polls left at their recorded
timestamps. This distinction is material: the live xApp ran with its
elevation state disabled, because the orchestration forwards only the environment
variables a container's compose block names and the three-dimensional flag was
not among them (Appendix~\ref{app:calib}). Equalising the estimator moves
wall-clock coverage from \SI{7.2}{\percent} to \MTcovWall\%: about half of the
published wall-clock coverage deficit was the filter, not the dilation.

On the wall clock not one of the \MTmotaTargetRuns{} target-runs is even mostly
tracked (\MTmotaWall); on the sensor's timeline \MTmotaSim{} are mostly, partly
and not tracked respectively, and fragmentation falls more than sixteenfold
(\MTfmWall{} to \MTfmSim{} per transit). We therefore do not claim that
multi-target tracking \emph{fails} on
this configuration. Nor do we claim it succeeds: \MTmotaSimLost{} of the
\MTmotaTargetRuns{} target-runs remain lost once
the artifact is removed, and that residue is the subject of the rest of this
section. What the dilation explains is the difference between those two
statements; what fails outright is the single-workstation twin's ability to
execute the scenario in real time, together with two evaluation constants that inherit
the resulting dilation.

\subsection{Effects Not Explained by Dilation}
\label{sec:cadence:residual}

The identifier surplus is not a cadence effect, and we report that as a
correction. The natural mechanism --- a track dying in a wall-clock detection gap
and being replaced, counting as a second identifier --- predicts that identifiers
per target should fall when the gap closes. Under one estimator it does not occur: identifiers
per target are \MTidsPerTgtWall{} on the wall clock and \MTidsPerTgtSim{} on the
sensor's timeline, a difference of no consequence. The apparent fall belonged to
the estimator change described above, not to the clock. The whole surplus is
therefore concurrent rather than sequential --- two identifiers held on one target
at the same time, which no change of timeline can explain --- and that is a
birth-admission problem, addressed in Section~\ref{sec:assoc:birth} and not yet
closed.

The weakest target likewise remains weakest, and the per-target spread is the
clearest signpost in this paper. On the sensor's timeline, and at the
\SI{20}{\metre} gate so that the two quantities are commensurable, UAV-A holds a
track for \MTaCovSim\% of its life, UAV-B \MTbCovSim\% and UAV-C \MTcCovSim\%.
Set beside the detection availability of the same three targets---\MTaPd\%,
\MTbPd\% and \MTcPd\% (Section~\ref{sec:assoc:availability})---the asymmetry is
the result: UAV-B and UAV-C are \emph{detected} far more often than they are
\emph{tracked}.
A target whose own detection is available in roughly three CPIs out of four, but
which holds a track in fewer than one in seven, is not detection-limited.
Whatever consumes that
margin lies between the detector and the track, and Section~\ref{sec:assoc:birth}
identifies admission as the mechanism. UAV-C's grazing geometry compounds it---the
same geometry costs it altitude observability
(Section~\ref{sec:measmodel:elcorr})---but geometry alone does not account for a
gap this size.

\section{Counter-UAS C2 Export Design}
\label{sec:c2}

The pipeline so far terminates at a visualization dashboard. An operational
deployment would deliver the tracks to a C2 system that
fuses heterogeneous sensors and cues effectors, and doing so without bespoke
per-vendor integration requires an \emph{open} sensor-to-C2 interface. This
section describes how the O-RAN sensing plane is made to speak one --- the SAPIENT
autonomous-sensor standard (BSI~Flex 335 / STANREC~4869) --- identifies the
track-quality prerequisite it imposes, and reports the continuity result that
meets it. What follows is a design, a first validation against an in-repository
mock fusion node, and a measured readiness result; live integration against a
commercial C2 remains future work.

\subsection{Adapter Architecture}
\label{sec:c2_adapter}

In SAPIENT terms the O-RAN sensing plane is an \emph{autonomous sensor module}
(ASM): an edge node that self-registers, advertises its coverage, and streams
detections and tracks to a fusion node (the high-level decision-making module,
HLDMM) over a length-prefixed protobuf stream. An adapter service subscribes to
the xApp's track output and maps it onto the standard message set: a
\emph{registration} message declaring the sensor's modality, location and
coverage volume; periodic \emph{detection} and \emph{track} messages carrying
target state; and \emph{status} heartbeats. The adapter is a thin, stateless
translation layer --- it holds no tracking logic of its own --- so the sensing
pipeline and the C2 interface evolve independently.

Two properties of a cellular ISAC sensor map naturally onto the standard. Its
coverage is not a fixed cone but a bistatic footprint that moves with the
scheduled UEs, which the registration/status coverage-volume field is designed to
carry as a time-varying report. And its native output is a range and bearing
relative to the array phase center, which the standard's range--bearing
representation carries directly; a geodetic position is an opt-in that requires a
surveyed sensor origin.

\subsection{Track Continuity as an Enabling Prerequisite}
\label{sec:c2_continuity}

A C2 consuming this feed needs \emph{stable} tracks: a single transiting UAV
should present as one track with a persistent identifier, not a stutter of
fragments. This is a track-management requirement distinct from raw detection ---
the front end detects in nearly every observable CPI, but detections arrive
intermittently (median gap \SI{\CAMPflatDetGapMed}{\second}, with dropouts up to
\SI{\CAMPflatDetGapMax}{\second}), and a purely detection-driven tracker goes
track-less between them.

We close this gap with a time-driven track-maintenance pass
(Section~\ref{sec:frag}): a low-rate loop forward-extrapolates each confirmed
track's posterior between detections and emits the predicted state, flagged as
coasted, bridging gaps up to a bounded horizon while keeping the track alive for
re-association under the same identifier. Over $N=\CAMPflatN$ seeded transits this
raises track continuity from \CAMPflatCont\% to \CAMPflatContBridge\%
(Section~\ref{sec:frag}), with no regression in measurement accuracy. Continuity,
not single-identity tracking, is what bridging buys: the transit is still covered
by \CAMPflatTrackIdsBridge{} identifiers per run (\CAMPflatTrackIds{} without
bridging), so a fusion node downstream must be tolerant of identifier churn ---
addressed in Section~\ref{sec:c2_roadmap}.

\paragraph{The continuity gain does not survive the sensor timeline}
The two figures above are scored on the \emph{wall clock}, which for this testbed
is not the timeline the scene evolves on (Section~\ref{sec:cadence}). Re-scoring
the same recorded detections at the sensor's own cadence --- same code, same five
seeds, only \texttt{ISAC\_COAST\_BRIDGE} and the timeline varying --- gives
\CAMPflatContSim\% unbridged against \CAMPflatContBridgeSim\% bridged. Bridging
buys \CAMPflatContSimDelta~percentage points there, against
\CAMPflatContWallDelta{} on the wall clock. The control that licenses reading
those numbers is that the same procedure reproduces both wall-clock figures above
to the decimal, spreads included.

The mechanism is implemented correctly; what it repairs is the emulator's
dilation. Single-target detections arrive
\SI{\CAMPflatDetGapMed}{\second} apart in scene time against a published-key
lifetime of \SI{\MTttl}{\second}, so no key expires until the wall clock stretches
that gap past it. The continuity prerequisite is met for a single target either way, but by the detector's cadence rather than by track maintenance, and the bridged wall-clock figure is in fact \emph{below} the
unbridged sensor-timeline one. We report both, and we take the sensor-timeline reading as the one a deployment should be
judged on.

\paragraph{Two conditions on the continuity figure}
The bridged continuity above is measured on the \emph{single-target} campaign of
the companion work, and the same thing happens here for the same reason: the gap
it closes does not exist on the sensor's timeline in either scenario. Enabling it
on the
\MTnTargets-target scene, over the same \MTcampN{} runs and detections, changes
track-level continuity by \MTcoastDelta~percentage points ($p=\MTcoastP$): every
record the bridge produces was already present, because the front end detects in
\MTcoastDetCov\% of intervals spaced \SI{\MTcpiPeriod}{\second} apart, so the
published track key never expires between them. Coasting can then only change a
record's provenance from measurement to extrapolation, and forcing it to fire
moves detection-level continuity from \MTcoastDetOff\% to \MTcoastDetOn\% while
leaving the track-level figure unchanged.

Bridging is a wall-clock remedy for a wall-clock gap, and on the sensor timeline
there is no such gap. We therefore report the multi-target continuity unbridged
throughout, and the readiness gate as met for a single target and \emph{not}
established for a multi-target feed --- the residual there is the per-target
geometry limit of Section~\ref{sec:assoc:birth}, which no track-maintenance
policy addresses.

Second, the inter-detection gaps above are on the sensor's timeline. A C2 consumes
the feed on the wall clock, and this testbed executes the scene about
\MTdilation$\times$ slower than it evolves (Section~\ref{sec:cadence}), so a
fusion node attached to a live run would see a track update roughly every
\SI{\MTwallGap}{\second} against a published-key lifetime of
\SI{\MTttl}{\second}. That is a property of the emulation rate, not of the sensor
or the tracker. Every multi-target number in this paper is scored on the sensor
timeline and is therefore unaffected; the two exceptions are the imported
single-target continuity figures above, which is why they are given on both
timelines. It does mean a live demonstration must either run faster than
real time or state the dilation explicitly.

\begin{figure*}[!tbp]
  \centering
  \includegraphics[width=\textwidth]{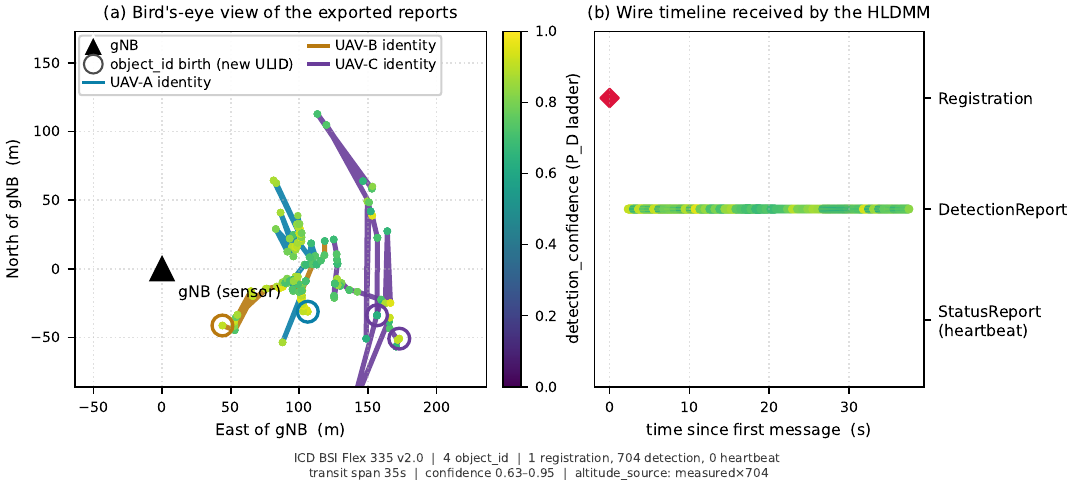}
  \caption{\textbf{Messages received by the counter-UAS C2} over one \MTnTargets-target transit, reconstructed from the wire feed (right-hand end of Fig.~\ref{fig:pipeline}, same run as Fig.~\ref{fig:track_overlay}). \emph{(a)} Exported reports around the gNB from their range--bearing field, colored by object identity, with detection confidence on the intensity scale. \emph{(b)} Message timeline: one registration, then a continuous report stream. The \MTnTargets{} targets arrive as distinct objects but under \emph{four} identities (UAV-C is handed over twice). Counts are this run's, not rates.}
  \label{fig:sapient_feed}
\end{figure*}

\subsection{Detection Confidence}
\label{sec:c2_confidence}

The standard's detection and track messages carry a confidence value that
downstream threat evaluation weights. Rather than a fixed nominal, we derive it
from the calibrated detection characterization of Section~\ref{sec:detchar}: the
measured probability of detection against echo-cell SNR --- and hence, through the
link budget, against cross section and bistatic range --- maps a detection's
reported cell SNR to a confidence. The exported value is therefore tied to the
sensor's measured physics rather than to a hand-set constant.

\paragraph{Detector-cell confidence versus track confidence}
The distinction is easy to lose across an interface and consequential when it is
lost, so we state it as a property of the exported field. The value answers:
\emph{given a detection in this range--Doppler cell at this SNR, what is the
probability that a target of the characterized cross-section band is present
there?} It is calibrated from the \emph{single-target} detector
characterization, on a scene with one target and no association step. It does
\emph{not} answer, and must not be read as answering, either of the two
questions a C2 actually weights: whether \emph{this track} corresponds to a real
object over its lifetime, and whether \emph{this identifier} has been carried by
the same object since it was minted. Three multi-target effects sit between the
cell and the track and none is in the calibration: a measured fraction of
reported detections correspond to no target (Section~\ref{sec:mt:rawcap}), two
targets share one detection in \MTcontested\% of intervals so a
high-confidence cell may be the \emph{wrong} target's
(Section~\ref{sec:assoc:availability}), and identifiers fragment
(Section~\ref{sec:assoc:results}) so confidence accumulated under one identifier
does not transfer to its successor. A fusion node that integrates this field
over a track's lifetime as though it were track confidence will therefore be
over-confident, and most over-confident exactly where contention is highest.

We report the field as \emph{detector-cell confidence} and do not export a track
or identity confidence, because neither is calibrated here. Calibrating them is
a specific piece of future work with a specific requirement: a reliability
diagram built under the multi-target false-detection load, on a scene where the
true association is known, so that a stated confidence can be checked against a
realized hit rate per track rather than per cell.

\subsection{Validation Against a Mock Fusion Node}
\label{sec:c2_validation}

Three separate exercises validate the export and they establish different
things; Table~\ref{tab:c2:validation} separates them, because quoting a result
from one while a reader assumes another is the way this subsection would
mislead. All three are \emph{schema-level validation against an in-repository
mock fusion node}. None is an interoperability result: no reference SAPIENT
middleware and no commercial C2 has ingested this feed.

\begin{table}[t]
  \caption{The three C2 export exercises, and what each does and does not
  establish. All are schema-level validation against an in-repository mock
  fusion node; none involves an external or commercial C2.}
  \label{tab:c2:validation}
  \footnotesize
  \setlength{\tabcolsep}{3.5pt}
  \begin{tabular}{@{}p{0.29\linewidth}p{0.30\linewidth}p{0.33\linewidth}@{}}
    \toprule
    Exercise & Establishes & Does not establish \\
    \midrule
    \textbf{Live single-target
    capture.} Adapter run as a containerized ASM beside the live stack;
    \num{110} detection reports reconstructed from the received bytes.
    Height prior, \emph{no} measured altitude.
    & Registration, message flow, status heartbeats and field mapping over a
      real socket, on a live run.
    & Anything multi-target; anything about the planar array's altitude. \\
    \addlinespace
    \textbf{Constructed \MTnTargets-track test.} The same adapter chain driven
    with \MTnTargets{} concurrent tracks.
    & That concurrent targets export as \emph{distinct}, stable objects and are
      never merged into one contact.
    & Behavior on real tracker output, which fragments. \\
    \addlinespace
    \textbf{Replayed multi-target feed} (Fig.~\ref{fig:sapient_feed}). Real
    adapter and real wire codec, driven offline from the sensor-timeline replay.
    Reports carry a \emph{measured} altitude.
    & What a C2 receives from tracker output as it actually is: four identities
      for \MTnTargets{} aircraft, guardrails suppressing the unconfirmed slot.
    & A live socket end to end; the replay is offline through the same codec. \\
    \bottomrule
  \end{tabular}
\end{table}

The first stage of the deployment path (Section~\ref{sec:c2_roadmap}) is now met.
Running the adapter as a containerized ASM alongside the sensing stack, it
self-registered with an in-repository mock fusion node and streamed the standard
message set over a length-prefixed protobuf socket for the duration of a live UPA
transit, reconstructed from exactly the bytes the fusion node received: a single
registration, \num{110}~detection reports, and periodic status heartbeats. That
capture is \emph{single-target}, and every figure in this subsection is from it;
the feed plotted in
Fig.~\ref{fig:sapient_feed} is the multi-target one described next.

\paragraph{Multiple simultaneous tracks}
A counter-UAS fusion node consumes a \emph{set} of tracks, so the property that
matters for this paper is that concurrent targets arrive as distinct, stable
objects rather than as one merged contact. Driving the same adapter chain --
track source, identity map, field mapping and protobuf codec -- with
\MTnTargets{} concurrent tracks yields \MTnTargets{} distinct object identifiers,
each a stable 26-character universally unique lexicographically sortable
identifier (ULID) that persists for as long as its track persists
and is re-minted only when the track disappears.

Fig.~\ref{fig:sapient_feed} is the end-to-end version of that check, and it is
the weaker claim of the two: rather than \MTnTargets{} tracks constructed for the
test, it replays the tracks a real transit actually produced. There the tracker
itself carried the \MTnTargets{} targets on five slots, of which four reached
confirmation; the adapter exported those four and suppressed the fifth, which
never left the tentative state --- so three UAVs reached the C2 as four objects.
The multiplicity is the tracker's and not the adapter's, which is the
distinction the two exercises together establish: nothing merges, the guardrails
withhold what was never confirmed, and what a fusion node has to absorb is
whatever fragmentation the tracker hands it. An identity is re-minted whenever a
track disappears and returns, and a re-mint is indistinguishable at the interface
from a new object --- which is why identity stability rather than detection is
the bottleneck for this export, and why the identity metrics of
Section~\ref{sec:assoc:results} are the ones that gate it.
Because these tracks carry a
measured altitude rather than the height prior, each report is additionally
labeled \texttt{altitude\_source=measured}: a consumer can tell an estimated
altitude from an assumed one, which is the operational point of delivering 3-D
tracks at all. Ground-truth and trail records, which share the same datastore,
are never exported under any of these conditions.

The field mapping behaves as specified in the single-target live capture above.
The reported range is
the gNB$\rightarrow$target slant range (spanning \SIrange{97}{231}{\metre}),
\emph{not} the roughly twofold-longer bistatic path the estimator carries
internally, so a C2 would place the target at its true stand-off rather than
twice as far. Confidence spans \numrange{0.35}{0.76}, tracking per-CPI SNR
through the calibrated ladder of Section~\ref{sec:c2_confidence} rather than a
constant. The honesty guardrails held: tentative and non-real tracks were
suppressed, no report preceded the first real detection, and the exporter reads
only track keys, never the simulator's ground-truth state.

That live reconstruction also makes the present limitations legible. Some
exported identifiers are clutter or false confirmed tracks rather than the true
transit, and the single UAV is fragmented across several identifiers --- the
adapter fails toward fragmentation rather than risk merging two UAVs. Both
are consistent with the track-management behavior of
Section~\ref{sec:c2_continuity} and motivate an export-side plausibility gate
(restricting to the illuminated volume with range-consistent confirmation) as the
next hardening step. In that capture, which predates the planar array, elevation
is still the configured height prior and the node advertises no altitude
observability --- unlike the
multi-target chain above, whose reports carry a measured altitude. Folding the
observability of Sections~\ref{sec:upa}--\ref{sec:multistatic} into the live
exported message is the immediate follow-on.

\subsection{Path to Deployment}
\label{sec:c2_roadmap}

The intended validation path is staged: first against an in-repository mock
fusion node that exercises the registration/detection/track/status message flow
and schema conformance, then against a reference SAPIENT middleware, and finally
against a commercial counter-UAS C2. The altitude observability of
Sections~\ref{sec:upa}--\ref{sec:multistatic} is what makes the geodetic-position
path meaningful: with height sensed rather than assumed, a surveyed sensor origin
yields a genuine 3-D track for the C2 rather than a 2-D track pinned to a nominal
altitude. Live integration against an external C2 --- and the interoperability and
latency testing it entails --- is the decisive next step and is left as future
work.

\section{Future Work}
\label{sec:future}

\emph{Multi-target identity.} The identity surplus is concurrent births, and Section~\ref{sec:assoc:ablation} measures the two candidate remedies as geometry-dependent trades rather than fixes: joint probabilistic association acts on coverage and localization, not identity, and widening the birth-inhibition region trades coverage for identity in the opposite direction. What neither ablation reaches is multi-hypothesis tracking, which defers the association decision instead of averaging over it. Beyond that, the scenario ends at \MTnTargets{} targets deliberately, and a swarm is a different problem rather than a larger one: co-moving targets share range and Doppler by construction, so the contention this paper measures at \MTcontested\% would approach unity, and the front end would need to resolve within a cell rather than pick between cells --- spatial smoothing or subarray averaging for a rank-greater-than-one covariance (Section~\ref{sec:mt:elevation}), and a group-tracking formulation that carries a formation as one object with an internal structure instead of $N_T$ competing tracks.

\emph{Angle accuracy and aperture.} The per-target position errors are governed by where each target crosses the array (Section~\ref{sec:assoc:birth}): an off-broadside transit degrades in cross-range as the single-snapshot interferometric estimator loses effective aperture toward endfire, and a low, close transit loses elevation observability. A subspace angle estimator with spatial smoothing would reduce the near-endfire bias, and a wider azimuth aperture --- more receive columns, or a multi-gNB deployment --- would hold cross-range precision across the larger angular excursions; the second bistatic pair of Section~\ref{sec:multistatic} moves the elevation boundary on the same principle.

\emph{Deployment-grade integration.} Section~\ref{sec:c2} makes the tracks
consumable by an external counter-UAS C2 over a standards-based adapter and
delivers the continuity prerequisite for a single target. What a deployment needs
next is the integration itself: the adapter exercised against a reference SAPIENT
middleware and then a commercial C2, with the interoperability and latency
testing that entails, and a surveyed sensor origin so the export carries a
geodetic 3-D position rather than a range--bearing report. The same track feed
has a second consumer with different requirements --- uncrewed-aircraft-system
(UAS) traffic management,
where the question is not threat evaluation but whether an observed track
reconciles with a filed flight plan and a broadcast remote identifier. That is a
correlation problem between a cooperative identifier and an uncooperative
measurement, it is where a network-native sensor should have an advantage over a
bolt-on radar. Both consumers also imply
duty-of-care work this evaluation does not address: a false-track rate stated as
an operational figure rather than a set-metric term, and a stated confidence a
downstream operator can act on.

\emph{Scenario coverage.} The evaluation uses one deterministic trajectory and
one gNB siting under $N=\MTcampN$ noise realizations, with a single alternative
heading as the only geometric variation. Multiple trajectories, speeds,
altitudes, cross sections and gNB mounting heights are needed before any
magnitude reported here generalizes. The per-target results make the reason
concrete: accuracy is governed by where each target flies relative to the array,
so a scenario is a sample of one siting, not of a capability.

\emph{Channel realism and hardware.} A richer multipath rendering --- more static
taps, and the UAV as an extended scatterer rather than a single collapsed echo
--- would improve non-line-of-sight robustness, and rotor micro-Doppler is absent
altogether, which is what a classification stage would need. Hardware-in-the-loop
validation with software-defined radios and a physical UAV remains the decisive
test, and it is the one that would convert every cadence figure in this paper
from a design rate into a measurement.

\section{Conclusion}
\label{sec:conclusion}

We evaluated multi-target UAV tracking end to end on an O-RAN ISAC simulation
testbed --- an OAI gNB with a UL-SRS PHY sensing stage, a FlexRIC Near-RT RIC
hosting an EKF xApp, and Sionna RT channel emulation --- against what a
counter-UAS command-and-control consumer requires rather than against detection
alone. On a single bistatic pair with a planar array, \MTnTargets{} UAVs are all
detected (per-target availability \MTaPd\%, \MTbPd\%, \MTcPd\%), yet the feed is
not usable by that consumer. The binding constraint is contention, not
sensitivity: two targets share one nearest detection in \MTcontested\% of
intervals. Two wall-clock constants in the evaluation path impose a coverage
ceiling of \MTceilingWall\% that re-scoring on the sensor timeline lifts
\MTcovWall\% to \MTcovSim\%. The vertical aperture buys association rather than
localization, and per-target accuracy (\SI{\MTaPosRmse}{\metre},
\SI{\MTbPosRmse}{\metre}, \SI{\MTcPosRmse}{\metre}) is a siting result, with only
the well-sited target inside the \SIrange{\MTkpiLo}{\MTkpiHi}{\metre} range
identified for the UAV use case in the 3GPP feasibility study. Concurrent tracks
are exported to a counter-UAS C2 over a standards-based sensor-to-C2 adapter
carrying a calibrated detection-confidence model, and that consumer's
requirements force the paper's sharpest configuration choice: a mechanism raising
coverage by \MTescCovDelta~percentage points --- measured on a scene
\MTelEnrichLo--\MTelEnrichHi$\times$ enriched in the very geometry it exploits,
so that gain is an upper bound --- is disabled because it costs identity at an
exchange rate no threshold setting improves.

The evaluation is $N=\MTcampN$ noise seeds on one deterministic trajectory and
one geometry, in emulation with no radios; the findings transfer as
\emph{mechanisms}. Hardware-in-the-loop
validation with software-defined radios and a live C2 integration trial remain
the decisive tests.

\appendices

\section{Material for Standalone Verification}
\label{app:standalone}

This appendix restates the minimum from the companion
manuscript~\cite{gurung_isac_testbed}: the
observation model the filter actually uses, the calibration constants and their
status, a summary of the detector characterization the exported confidence is
derived from, the clutter-rank choice, and where the code and data are.

\subsection{Observation Model}
\label{app:obs}

The state is $\mathbf{x} = [x, y, z, v_x, v_y, v_z]^{\mathsf{T}}$ in the ENU
frame, with the gNB array at $\mathbf{p}_G$ and the transmitting nrUE at
$\mathbf{p}_U$ separated by the known baseline $L = \lVert \mathbf{p}_U -
\mathbf{p}_G \rVert$. Writing $\mathbf{p} = [x,y,z]^{\mathsf{T}}$,
$\mathbf{d}_G = \mathbf{p} - \mathbf{p}_G$ and $\mathbf{d}_U = \mathbf{p} -
\mathbf{p}_U$, the four measurements are
\begin{align}
  h_R(\mathbf{x})    &= \lVert \mathbf{d}_G \rVert + \lVert \mathbf{d}_U \rVert,
  \label{eq:app:range}\\
  h_v(\mathbf{x})    &= \mathbf{v}^{\mathsf{T}}
                        \!\left( \frac{\mathbf{d}_G}{\lVert \mathbf{d}_G \rVert}
                               + \frac{\mathbf{d}_U}{\lVert \mathbf{d}_U \rVert}
                        \right),
  \label{eq:app:dop}\\
  h_{\mathrm{az}}(\mathbf{x}) &= \arcsin\!\left(
        \frac{d_{G,y}}{\lVert \mathbf{d}_G \rVert} \right),
  \label{eq:app:az}\\
  h_{\mathrm{el}}(\mathbf{x}) &= \arcsin\!\left(
        \frac{d_{G,z}}{\lVert \mathbf{d}_G \rVert} \right).
  \label{eq:app:el}
\end{align}
First, \eqref{eq:app:range} is the \emph{total} bistatic
path; the detector reports the excess path $\Delta R_{\mathrm{bi}} = h_R - L$
relative to the direct arrival, and the tracker restores $h_R$ by adding the known
$L$. Mixing the two conventions is a \SI{130}{\metre} error on this geometry.
Second, \eqref{eq:app:az} is a \emph{direction cosine} normalized by the
three-dimensional range $\lVert \mathbf{d}_G \rVert$, not the ground-plane bearing
$\operatorname{atan2}(d_{G,y}, d_{G,x})$; the two differ by
$\cos(\text{elevation})$ and the consequences of confusing them are measured in
Section~\ref{sec:measmodel:cone}. Third, both angles are formed at the gNB
aperture only --- the nrUE contributes delay and Doppler but no bearing --- which
is why a single bistatic pair with one array has the observability structure
Section~\ref{sec:mt:altitude} reports.

The measurement covariance is diagonal,
$\mathbf{R} = \operatorname{diag}(\sigma_R^2, \sigma_v^2,
\sigma_{\mathrm{az}}^2, \sigma_{\mathrm{el}}^2)$, with the values of
Table~\ref{tab:params}. They are set from the \emph{outlier-inclusive}
truth-referenced residual spread rather than from a robust estimate: the residual
distribution is a mixture, and a robust fit describes its core while the filter
must survive its tail.

\subsection{Calibration Constants and Their Status}
\label{app:calib}

Four constants are calibrations rather than derived quantities.

\begin{enumerate}
  \item \textbf{Uplink receive gain}, \SI{\MTgainUl}{\decibel}, fixed. Set to the
        value the emulator's own headroom-safe controller settles at
        (\SI{\MTgainSettle}{\decibel}) rather than left adaptive, because the
        controller ramps between SRS bursts and over-drives the receiver. In
        force; the clipped-block count is the validity check.
  \item \textbf{Bistatic range bias}, \SI{\MTrangeBias}{\metre}.
        \emph{Characterized but not in force} --- see
        Section~\ref{sec:measmodel} and
        Table~\ref{tab:measmodel:ablation}. Every absolute range figure in this
        paper carries the uncorrected offset.
  \item \textbf{Angular and range measurement standard deviations},
        \MTsigmaAzMeas\si{\degree} and \SI{\MTsigmaRMeas}{\metre}, re-estimated
        against ground truth after the cone-angle correction. In force.
  \item \textbf{Altitude variance floor}, constraining the reported altitude
        variance to what one elevation look supports
        (Section~\ref{sec:measmodel:elcorr}). In force.
\end{enumerate}

\subsection{Detector Characterization}
\label{app:detchar}

The exported detection confidence (Section~\ref{sec:c2_confidence}) is a mapping
from reported cell SNR to $P_{\mathrm{D}}$, and is only as good as the
characterization behind it. That characterization is a single-target echo-SNR
sweep --- \CAMPpdLevels{} cross-section levels spanning \CAMPpdRcsSpan~dB,
\CAMPpdSeeds{} seeds, \CAMPpdCpis{} CPIs --- summarized as follows. At the
operating point ($\CAMPpdOpSnr$~dB cell SNR) $P_{\mathrm{D}} = \CAMPpdOpPd$.
Across the swept band $P_{\mathrm{D}}$ runs from \CAMPpdRcsLoPd{} at
$\CAMPpdRcsLo$~dBsm to \CAMPpdRcsHiPd{} at $\CAMPpdRcsHi$~dBsm, and saturates
above $\CAMPpdSatSnr$~dB. The floor is \CAMPpdFloorPd{} and is whole-CPI dropout
rather than per-cell miss. Realized false-alarm probability at the deployed
threshold is $\CAMPpfaAfter$, against a nominal $10^{-4}$.

Two limitations of this curve carry into the exported confidence and are the
substance of Section~\ref{sec:c2_confidence}: it is measured with one target
present, so it contains no association ambiguity and no multi-target false-alarm
load; and it is a per-cell quantity, so it does not support a statement about a
track.

\subsection{Clutter-Subspace Rank}
\label{app:eca}

The deflation stage projects out the leading $K_c$ eigenvectors of the clutter
covariance before detection. $K_c = 1$ is a correctness requirement rather than a tuning choice on this scene: in a flat line-of-sight geometry the direct
path is a single dominant component and the UAV echo is the second, so $K_c \ge
2$ removes the target with the clutter. This is the single most consequential
constant in the sensing chain --- at $K_c = 2$ the detector reports the scene as
empty --- and it is stated here because a reader cannot otherwise tell an
empty-scene result from a canceled-target one.

\subsection{Code and Data Availability}
\label{app:data}

Every number in this paper is a macro generated from a committed artifact rather
than typed by hand. The paper directory carries: the aggregated campaign outputs
for both arms of every comparison; the pinned single-seed run (detections, gNB
log and manifest) from which the figures are regenerated; a single script that
rebuilds all figures from that run through the same replay and association the
tables are scored from; the analysis scripts behind each provenance note,
including the ones that overturned earlier claims of ours; and per-campaign
manifests recording image digests, repository revision and trace checksums.

\section*{Acknowledgements}
The authors would like to thank the OpenAirInterface (OAI),
the FlexRIC, and NVIDIA Sionna RT teams for the open-source projects.
Also we would like to thank Google Cloud for providing the CPU/GPU resources to run the experiments, and
GitHub for providing the repository hosting.

\section*{Declaration of Generative AI Use}
During the preparation of this work the authors used Anthropic Claude (including
Claude Code) and OpenAI ChatGPT for software implementation and testing of the
testbed, for data post-processing and analysis scripting, and for language
editing and manuscript refinement. The authors reviewed and edited the content as
needed and take full responsibility for the content of the published article. No
generative-AI tool is listed as an author, and none was used to generate,
fabricate or alter measurement data: every quantitative result in this paper is
produced by the instrumented pipeline described herein from committed campaign
artifacts.

\bibliographystyle{IEEEtran}
\bibliography{references}

@techreport{3gpp_tr38867,
  author      = {{3GPP}},
  title       = {{Study on Channel Model for NR Integrated Sensing
                 and Communication}},
  institution = {3rd Generation Partnership Project (3GPP)},
  type        = {Technical Report},
  number      = {TR 38.867},
  year        = {2025},
  note        = {Release-19 study item; study report, not a normative specification}
}

@techreport{3gpp_tr22837,
  author      = {{3GPP}},
  title       = {{Feasibility Study on Integrated Sensing and Communication}},
  institution = {3rd Generation Partnership Project (3GPP)},
  type        = {Technical Report (Study Item)},
  number      = {TR 22.837},
  year        = {2024},
  note        = {V19.4.0; study report, not a normative specification}
}

@misc{3gpp_cr38901_isac,
  author       = {{3GPP}},
  title        = {{CR to introduce channel model for Integrated Sensing and Communication}},
  howpublished = {Change Request CR~38.901-0027, Release~19, TSG status: approved},
  year         = {2025},
  note         = {3GPP Portal, CrId~577179; introduces an ISAC channel model into TR~38.901.
                  Channel-model work, not a normative sensing waveform or service specification}
}

@techreport{oran_wg1_arch,
  author      = {{O-RAN Alliance}},
  title       = {{O-RAN Architecture Description}},
  institution = {O-RAN Alliance},
  type        = {Technical Specification},
  number      = {O-RAN.WG1.O-RAN-Architecture-Description-v07.00},
  year        = {2022}
}

@techreport{oran_wg3_e2ap,
  author      = {{O-RAN Alliance}},
  title       = {{O-RAN E2 Application Protocol (E2AP)}},
  institution = {O-RAN Alliance},
  type        = {Technical Specification},
  number      = {O-RAN.WG3.E2AP-R003},
  year        = {2024},
  note        = {Defines the RIC subscription/indication procedures an E2
                 service model must comply with}
}

@article{polese_dapp,
  author  = {Polese, Michele and Bonati, Leonardo and D'Oro, Salvatore
             and Basagni, Stefano and Melodia, Tommaso},
  title   = {{dApps}: Distributed Applications for Real-Time Inference
             and Control in {O-RAN}},
  journal = {arXiv preprint arXiv:2203.02370},
  year    = {2022}
}

@article{oran_isac_dapp,
  author  = {Eduardo Baena and Rajesh Krishnan and Mai Vu and Gil Zussman and Dimitrios Koutsonikolas},
  title   = {Toward Native {ISAC} Support in {O-RAN} Architectures for {6G}},
  journal = {arXiv preprint arXiv:2603.03607},
  year    = {2026}
}

@techreport{ngrg_dapp,
  author      = {{O-RAN Alliance nGRG}},
  title       = {{dApp} Use Cases and Requirements},
  number      = {nGRG-RR-2024-10},
  institution = {O-RAN Alliance},
  year        = {2024}
}

@article{liu_survey_isac,
  author  = {F. Liu and C. Masouros and A. P. Petropulu and
             H. Griffiths and L. Hanzo},
  title   = {{Joint Radar and Communication Design:
             Applications, State-of-the-Art, and the Road Ahead}},
  journal = {IEEE Trans. Commun.},
  volume  = {68},
  number  = {6},
  pages   = {3834--3862},
  year    = {2020},
  doi     = {10.1109/TCOMM.2020.2973976}
}

@article{zhang_dual_function,
  author  = {F. Liu and L. Zheng and Y. Cui and C. Masouros and
             A. P. Petropulu and H. Griffiths and Y. C. Eldar},
  title   = {{Integrated Sensing and Communications: Towards
             Dual-Functional Radar-Communication Systems}},
  journal = {IEEE J. Sel. Areas Commun.},
  volume  = {40},
  number  = {6},
  pages   = {1728--1747},
  year    = {2022},
  doi     = {10.1109/JSAC.2022.3156632}
}

@inproceedings{kaltenberger_oai,
  author    = {F. Kaltenberger and A. Khodakarami and C. De Vleeschouwer
               and L. S. Cardoso and R. Knopp},
  title     = {{OpenAirInterface: An Open-Source Software Radio
               Platform for 5G Research}},
  booktitle = {Proc. IETF/IRTF Open Networking Summit},
  year      = {2020}
}

@inproceedings{schmidt_flexric,
  author    = {R. Schmidt and S. Shariat and A. Karimzadeh and
               F. Kaltenberger and A. Clemente and N. Nikaein},
  title     = {{FlexRIC: An SDK for Next-Generation Disaggregated
               and Programmable RANs}},
  booktitle = {Proc. ACM CoNEXT},
  year      = {2021},
  doi       = {10.1145/3485983.3494870}
}

@article{hoydis_sionna,
  author        = {J. Hoydis and S. Cammerer and F. A. Aoudia and
                   A. Vem and N. Binder and G. Marcus and A. Keller},
  title         = {{Sionna: An Open-Source Library for Next-Generation
                   Physical Layer Research}},
  journal       = {arXiv preprint},
  year          = {2022},
  eprint        = {2203.11854},
  archiveprefix = {arXiv},
  primaryclass  = {cs.IT},
  url           = {https://arxiv.org/abs/2203.11854}
}

@article{colone_eca_passive_radar,
  author  = {F. Colone and D. W. O'Hagan and P. Lombardo and C. J. Baker},
  title   = {{A Multistage Processing Algorithm for Disturbance Removal
             and Target Detection in Passive Bistatic Radar}},
  journal = {IEEE Trans. Aerosp. Electron. Syst.},
  volume  = {45},
  number  = {2},
  pages   = {698--722},
  year    = {2009},
  doi     = {10.1109/TAES.2009.5089551}
}

@article{roy_esprit,
  author  = {Roy, Richard and Kailath, Thomas},
  title   = {{ESPRIT} -- Estimation of Signal Parameters via Rotational
             Invariance Techniques},
  journal = {IEEE Trans. Acoust., Speech, Signal Process.},
  volume  = {37},
  number  = {7},
  pages   = {984--995},
  year    = {1989}
}

@article{rohling_oscfar,
  author  = {Rohling, Hermann},
  title   = {Radar {CFAR} Thresholding in Clutter and Multiple Target
             Situations},
  journal = {IEEE Trans. Aerosp. Electron. Syst.},
  volume  = {AES-19},
  number  = {4},
  pages   = {608--621},
  year    = {1983}
}

@article{bauhofer_mtt_isac,
  author  = {Bauhofer, Maximilian and Henninger, Marcus and Kottkamp, Meik
             and Giroto, Lucas and Grill, Philip and Felix, Alexander
             and Wild, Thorsten and ten Brink, Stephan and Mandelli, Silvio},
  title   = {Experimental Demonstration of Multi-Target Tracking in Integrated
             Sensing and Communication},
  journal = {arXiv preprint arXiv:2510.22180},
  year    = {2025}
}

@article{saur_reliable_uav,
  author  = {Saur, Stephan and Doll, Mark and Grudnitsky, Artjom
             and Mandelli, Silvio and Giroto, Lucas and Henninger, Marcus
             and Wild, Thorsten},
  title   = {Reliable {UAV} Detection with {ISAC}},
  journal = {arXiv preprint arXiv:2605.23561},
  year    = {2026}
}

@article{huang_fuse_then_detect,
  author  = {Huang, Wenyu and Gonz{\'a}lez-Prelcic, Nuria and Ratnam, Vishnu
             and Bayraktar, Murat and Zhang, Charlie Jianzhong},
  title   = {Fuse-then-Detect for Passive {UAV} Localization Using Multi-{UE}
             {5G} Uplink Signals},
  journal = {arXiv preprint arXiv:2607.11955},
  year    = {2026}
}

@article{varshney_multitrp_uav,
  author  = {Varshney, Neeraj and Blandino, Steve and Wang, Jian
             and Bodi, Anuraag and Gentile, Camillo and Golmie, Nada},
  title   = {Multi-{TRP} Assisted {UAV} Detection in {3GPP} {5G}-Advanced
             {ISAC} Network},
  journal = {arXiv preprint arXiv:2604.26113},
  year    = {2026}
}

@article{sagduyu_multiscout,
  author  = {Sagduyu, Yalin E. and Davaslioglu, Kemal and Erpek, Tugba
             and Kompella, Sastry and Anderson, Gustave and Ashdown, Jonathan},
  title   = {{MULTI-SCOUT}: Multistatic Integrated Sensing and Communications
             in {5G} and Beyond for Moving Target Detection, Positioning,
             and Tracking},
  journal = {arXiv preprint arXiv:2507.02613},
  year    = {2025}
}

@article{polese_dapps_6gr,
  author  = {Polese, Michele and Gangula, Rajeev and Melodia, Tommaso},
  title   = {Enabling Programmable Inference and {ISAC} at the {6GR} Edge
             with {dApps}},
  journal = {arXiv preprint arXiv:2603.29146},
  year    = {2026}
}

@misc{att_ericsson_drone,
  author       = {{AT\&T} and {Ericsson}},
  title        = {The Future Takes Flight: {AT\&T} and {Ericsson} Demonstrate
                  Drone Detection Outside of {AT\&T} Stadium},
  howpublished = {Press release},
  year         = {2026},
  month        = jul,
  note         = {\url{https://about.att.com/story/2026/att-ericsson-drone-detection.html}}
}

@misc{gurung_isac_testbed,
  author = {Gurung, Arun K. and Sathananthan, Satha K. and Pokhrel, Shiva R.},
  title  = {{5G {ISAC}-Based {UAV} Detection and {3-D} Tracking Using Uplink
            Sounding Reference Signals on an End-to-End {O-RAN} Simulation Testbed}},
  year   = {2026},
  note   = {arXiv:2608.05826 [cs.NI]}
}

\end{document}